\documentclass[aps,prd,nofootinbib,twocolumn,groupedaddress]{revtex4}
\usepackage{dcolumn}
\usepackage{bm}
\usepackage{pdfpages}
\usepackage{amsmath}    
\usepackage{amsfonts} 
\usepackage{amssymb}
\usepackage{mathptmx}      
\usepackage{graphicx}   
\usepackage{mathtools}
\usepackage{graphics}
\usepackage{epsfig}
\usepackage{tabularx}   
\usepackage{multirow}
\usepackage{bibentry}
\usepackage{braket}
\usepackage{algorithm}
\usepackage{algorithmic}
\usepackage{listings}
\usepackage{setspace}
\usepackage{fancyhdr}
\usepackage{slashed}
\usepackage{color}
\usepackage{soul}
\usepackage{xfrac}
\usepackage[pdftex, pdfborder= 0 0 0, citecolor=blue, urlcolor=blue, linkcolor=blue, colorlinks=true, bookmarksopen=true]{hyperref}
\def\Journal#1#2#3#4{{#1} {\bf#2}, #3 (#4)}

\def\NPA{{\rm Nucl. Phys.} A}
\def\NPB{{\rm Nucl. Phys.} B}
\def\PLB{{\rm Phys. Lett.}  B}
\def\PRL{\rm Phys. Rev. Lett.}
\def\PRD{{\rm Phys. Rev.} D}
\def\PRC{{\rm Phys. Rev.} C}
\def\ZPC{{\rm Z. Phys.} C}

\def\EPJC{{\rm Eur. Phys. J.} C}

\def\ep{\epsilon}
\def\vep{\varepsilon}
\def\la{\langle}
\def\ra{\rangle}

\def\lam{\lambda}
\def\al{\alpha}

\def\be{\begin{equation}}
\def\ee{\end{equation}}
\def\bea{\begin{eqnarray}}
\def\eea{\end{eqnarray}}
\begin{document}
\title{Chiral anomaly and the pion properties in the light-front quark model}
%
     
\author{ Ho-Meoyng Choi}
\affiliation{Department of Physics, Teachers College, Kyungpook National University,
     Daegu 41566, Korea}
     
 \author{Chueng-Ryong Ji}
\affiliation{Department of Physics, North Carolina State University,
Raleigh, North Carolina  27695-8202, USA} 
\begin{abstract}
{ We explore the link between the chiral symmetry of QCD and the numerical results of the light-front quark model, analyzing both the two-point and three-point functions of the pion. Including the axial-vector coupling as well as the pseudoscalar coupling in the light-front quark model, we discuss the implication of the chiral anomaly in describing the pion decay constant, the pion-photon transition form factor and the electromagnetic form factor of the pion. In constraining the model parameters, we find that the chiral anomaly plays a 
critical role and the analysis of $F_{\pi\gamma}(Q^2)$ in timelike region is important. Our results indicate that the constituent quark picture is effective for the low and high $Q^2$ ranges 
implementing the quark mass evolution effect as $Q^2$ grows.}
\end{abstract}
\maketitle

\section{Introduction}
\label{sec:I}
Due to just a single hadron involvement, 
the meson-photon transition is well known to be the simplest 
exclusive process in testing the quantum chromodynamics (QCD) 
and understanding the structure of the meson~\cite{Lepage:1980fj}.
As the pion is regarded as the lightest pseudo-Goldstone boson arising from the spontaneous symmetry breaking of the chiral symmetry in QCD, it is particularly important to analyze the pion production process via the two-photon collision, $\gamma^*\gamma\to \pi$,
which involves only one transition form factor (TFF) $F_{\pi\gamma}(Q^2)$, where $q^2=-Q^2$ is
the squared momentum transfer of the virtual photon. 
Its complete understanding requires a formulation capable of explaining both the nonperturbative Adler-Bell-Jackiw (ABJ) anomaly (or the chiral anomaly)~\cite{ABJ},
which determines $F_{\pi\gamma}(0)$ when both photons are real (i.e. $Q^2=0$), and simultaneously the perturbative QCD (pQCD) prediction that governs the behavior of  $F_{\pi\gamma}(Q^2)$ at large momentum transfer $Q^2$ region.
Since
the publication of the $BABAR$~\cite{Aubert:2009mc}
 for $F_{\pi\gamma}(Q^2)$ in 
$0\leq Q^2\leq 40$ GeV$^2$ showing the violation of the scaling law predicted by pQCD~\cite{Lepage:1980fj}, 
many theoretical efforts~\cite{MS09,Ra09,MP09,AB03,Dor10,DK13,Kroll11,Roberts:2010rn} 
have been made to clarify this issue.

In an effort to examine the issue of the scaling behavior of $Q^2 F_{\pi\gamma}(Q^2)$, we have attempted to analyze the corresponding form factor not only in the spacelike region but also in the timelike region~\cite{Choi:2017zxn}. 
In particular, we presented the new direct method to explore the timelike region without resorting to mere analytic continuation from spacelike to timelike region~\cite{Choi:2017zxn}. Our direct calculation in timelike region showed the complete agreement with not only the analytic continuation result from spacelike region but also the result from the dispersion relation (DR) between the real and imaginary parts of the form factor.
This development added more predictability to the light-front quark models 
(LFQMs)~\cite{Jaus91,Choi:1997iq,Choi:1999nu,cheng04,Choi:2007yu,Choi:2007se,WH10,deMelo:2013zza,LG10,Choi:2017zxn,CRJ19} which have been successful in describing hadron phenomenology
based on the constant constituent quark and antiquark masses. 
More specifically, our LFQM~\cite{Choi:1997iq,Choi:1999nu} built on the variational principle to the QCD-motivated Hamiltonian provided a good description of the pion electromagnetic and transition form factors~\cite{Choi:1997iq,Choi:2007yu,Choi:2017zxn,CRJ19}.

We have further discussed the link between the chiral symmetry of QCD and the numerical results of the LFQM, analyzing both the two-point and three-point functions of a pseudoscalar meson from the perspective of the vacuum fluctuation consistent with the chiral symmetry of QCD~\cite{CJ2015}.
This link is due to a pair creation of particles with zero light-front (LF) longitudinal momenta from the vacuum, which captures the vacuum effect for the consistency with the chiral symmetry properties of the strong interactions. With this link,
the zero-mode contribution~\cite{cheng04,Zero1,Zero2,Zero3,Jaus99,BCJ01,BCJ03}
in the meson decay process could effectively accommodate the effect of vacuum fluctuation consistent with the chiral symmetry of the strong interactions. In this respect, the LFQM with effective degrees of freedom represented by the constituent quark and antiquark was linked to the QCD since the zero-mode link to the QCD vacuum could provide the view of an effective zero-mode cloud around the quark and antiquark inside the meson. While the constituents are dressed by the zero-mode cloud, they are expected to satisfy the chiral symmetry consistent with the QCD. Our numerical results~\cite{CJ2014} were indeed consistent with this expectation and effectively indicated that the constituent quark and antiquark in the standard LFQM~\cite{Choi:2007yu,Choi:1997iq,Choi:1999nu,Choi:2007se,Choi:2017zxn,CRJ19} could be considered 
indeed as the dressed constituents including the zero-mode quantum fluctuations from the QCD vacuum.

Moreover, the lattice QCD results~\cite{Kuramashi} indicated that the mass difference between $\eta^\prime$ and pseudoscalar octet mesons due to the complicated nontrivial vacuum effect increases or decreases as the extrapolating quark mass decreases or increases; i.e., the effect of the topological charge contribution should be small as the quark mass increases. This correlation between the quark mass and the nontrivial QCD vacuum effect further supported the development of our LFQM~\cite{Choi:1997iq} because the complicated non-trivial vacuum effect in QCD could be traded off by rather large constituent quark masses.
As a precursor of this development of LFQM, the constituent quark model in the light-front quantization approach
appeared based on the spin-averaged mass scheme~\cite{Dziem,CJ90} of taking the $\pi$ and $\rho$ meson masses equal to the spin-averaged value $M_{\rm av}=(\frac{1}{4}M_\pi +\frac{3}{4}M_\rho)_{\rm Exp}\approx 612$ MeV.
In retrospect, such early development was an attempt to trade off the nonperturbative QCD effect with the constituent quark mass averaged between the $\pi$ and $\rho$ mesons although the spin-averaged mass scheme itself was too naive to accommodate the complicate non-trivial vacuum effect. 
More sophisticated analysis was developed later to take into account the effect of the mass evolution (from constituent to current quark mass) on $F_\pi(Q^2)$ at low and intermediate $Q^2$~\cite{KCJ01}. 
We have then also discussed a constraint of conformal symmetry in the analysis of the pion elastic form factor both in spacelike and timelike regions~\cite{ConformalCJ1,ConformalCJ2}, confirming the anti-de Sitter space 
geometry/conformal field theory (AdS/CFT) correspondence~\cite{AdS1}.

While the early LFQM approach of the spin-averaged mass scheme~\cite{Dziem, CJ90} included
both the pseudoscalar and axial-vector couplings for the pseudoscalar meson vertex, 
only the specific vertex given by $\Gamma_\pi=(M_\pi + {/\!\!\!\!\! P})\gamma_5$ 
with the four momentum $P^\mu$ was taken 
for the coupling with the constituent quark and antiquark in the triangle 
loop amplitude. Since then, the later development of most standard 
LFQM~\cite{Jaus91,Choi:1997iq,Choi:1999nu,cheng04,Choi:2007yu,Choi:2007se,WH10,deMelo:2013zza,LG10,Choi:2017zxn,CRJ19}
including ours~\cite{Choi:1997iq,Choi:2007yu,Choi:2017zxn} built on the variational principle used typically only the pseudoscalar vertex
given by $\Gamma_\pi = A_\pi \gamma_5$, where $A_\pi$ is a constant of proportionality with the mass dimension which
gets absorbed into the normalization of the spin-orbit wave function.
However, the generalization of the vertex including the axial vector coupling deserves 
further consideration to include the exact chiral limit $(M_\pi, m\to 0)$ phenomena, where $m$ represents the $u(d)$ quark mass respecting isospin symmetry.
In particular, the ABJ anomaly~\cite{ABJ} is the key to understand the $\pi^0 \to \gamma\gamma$ decay rate
resolving the issue with the Sutherland-Veltman theorem~\cite{Sutherland-Veltman}.
As the Thompson low-energy limit works for the Compton scattering on any target,
the Sutherland-Veltamn theorem reveals that the nonanomalous term must vanish in the case
when both photons are on-mass-shell~\cite{Pham}. Only the chiral anomaly is capable of 
explaining the $\pi^0$ decay to the two real photons. Thus, it appears important to analyze the 
contribution from the axial-vector coupling  together with the contribution from the pseudoscalar coupling
to explore the correlation between the nontrivial QCD vacuum effect and  
the constituent quark mass as well as the parameter of the trial wave function 
in the LFQM built on the variational principle.    

In this work, we include the axial-vector coupling in addition to the pseudoscalar coupling
in our LFQM for the pion to explore a well-defined chiral limit providing still a good description 
of the pion electromagnetic and transition form factors~\cite{Choi:1997iq,Choi:2007yu,Choi:2017zxn}.
To examine the relative contribution between the pseudoscalar coupling and
the axial-vector coupling, we take the more general vertex 
$\Gamma_\pi=(A _\pi + B_\pi {/\!\!\!\!\! P})\gamma_5$ 
which goes beyond the specific vertex $\Gamma_\pi=(M_\pi + {/\!\!\!\!\! P})\gamma_5$ 
previously taken in the spin-averaged mass scheme~\cite{Dziem, CJ90} for the pion spin-orbit structure with the four momentum $P^\mu$.
We then describe the pion properties such as $f_\pi, F_{\pi\gamma}(Q^2), F_{\pi}(Q^2)$
depending on the variation of the quark mass in a  self-consistent manner within this model.

The paper is organized as follows. In Sec.~\ref{sec:II}, we introduce the spin-orbit wave function of the pion
obtained from the operator $\Gamma_\pi=(A _\pi + B_\pi {/\!\!\!\!\! P})\gamma_5$ and show the chiral limit
expression of the spin-orbit wave function. We also compare it with our previous spin-orbit wave function
obtained from the operator $\Gamma_\pi= A_\pi \gamma_5$.
In Sec.~\ref{sec:III}, we apply our LFQM for the calculation of $f_\pi, F_{\pi\gamma}(Q^2)$ and $F_{\pi}(Q^2)$ using both
constituent quark mass and the chiral limit result. Especially, we explicitly
obtain the analytic form of $f_\pi$ and $F_{\pi\gamma}(Q^2)$ in the exact chiral limit ($M_\pi, m\to 0$).
We also show that  our 
chiral limit result for twist-2 pion distribution amplitude (DA), which encodes the nonperturbative information on the pion,
is exactly the same as the anti-de Sitter/conformal 
field theory (AdS/CFT) prediction of the asymptotic DA~\cite{AdS1,AdS2,AdS3}.
In Sec.~\ref{sec:IV}, we  discuss how to determine the model parameters and show the numerical results of the pion DA,
the pion TFF  $F_{\pi\gamma}(Q^2)$ both in spacelike and timelike regions covering the full momentum 
transfer region, and the pion electromagnetic form factor $F_{\pi}(Q^2)$ in the spacelike region. In this section, 
we show the results by varying the quark mass in a self-consistent way so that one can effectively 
see the evidence of quark mass evolution effect as $Q^2$ changes for $F_{\pi\gamma}(Q^2)$ and $F_{\pi}(Q^2)$. Summary and conclusions
follow in Sec.~\ref{sec:V}.
In the Appendix, we provide the derivation of our new spin-orbit wave function.

\section{Model Description}
\label{sec:II}
The key approximation in the LFQM is the mock-hadron approximation~\cite{Hayne} to saturate the Fock state expansion by the
constituent quark and antiquark and treat that Fock state as a free state as far as the spin-orbit part is concerned.
The assignment of the quantum numbers such as angular momentum, parity and charge conjugation to the LF wave
function is given by the Melosh transformation~\cite{Mel}.
For example, the pion state $|\pi\ra$ is represented by $|\pi\ra = \Psi^\pi_{Q\bar{Q}}|Q\bar{Q}\ra$, where $Q(\bar{Q})$
is the effective dressed quark (antiquark). That is, the pion state as a valence $Q\bar{Q}$ bound state with 
momentum $P^\mu = (P^+, P^-, {\bf P}_\perp)$ is determined by the
light-front  wave function (LFWF)  
%
\be\label{eq1}
\Psi^\pi_{Q\bar{Q}}\equiv\Psi_{\pi}(x_i,{\bf k}_{i\perp},\lambda_i)
=\phi_R(x_i,{\bf k}_{i\perp})\chi(x_i,{\bf k}_{i\perp},\lambda_i),
\ee
where $x_i = k^+_i/P^+$, ${\bf k}_{i\perp}$, and $\lambda_i$
are the Lorentz-invariant  longitudinal-, transverse-momenta and the helicity of each constituent quark (antiquark), respectively,
with the properties satisfying $\sum_{i=1}^{2} x_i =1$ and $\sum_{i=1}^{2} {\bf k}_{i\perp}=0$. 
Here, $\phi_R$ is the radial wave function 
which is taken as the trial wave function following the variational principle
and $\chi$ is
the LF spin-orbit wave function which is obtained by the interaction-independent Melosh transformation from the ordinary
equal-time static spin-orbit wave function 
assigned by the quantum numbers $J^{PC}$. 

The LFWF is normalized according to
\be\label{eq2}
\la\Psi^\pi_{Q\bar{Q}} | \Psi^\pi_{Q\bar{Q}}\ra = P_{Q\bar{Q}},
\ee
where $P_{Q\bar{Q}}$ is the probability of finding the $Q\bar{Q}$ component in the LFWF.
For the radial wave function $\phi _R(x,{\bf k}_\perp)$ of the pion 
with the same constituent quark and antiquark masses 
$m_Q=m_{\bar Q}\equiv m$, 
we 
take the harmonic oscillator (HO) wave function as our trial wave function
\be\label{eq3}
\phi_{R}(x,{\bf k}_{\perp}) = \sqrt{P_{Q\bar{Q}}}
\frac{4\pi^{3/4}}{\beta^{3/2}} 
\sqrt{\frac{\partial k_z}{\partial x}} e^{-\frac{{\vec k}^2}{2\beta^2}},
\ee
where
$\partial k_z/\partial x = M_0/4x(1-x)$  
is the Jacobian of the variable transformation
$\{x,{\bf k}_\perp\}\to {\vec k}=({\bf k}_\perp, k_z)$ with
 $M^2_0=({\bf k}^2_\perp + m^2)/x(1-x)$ being the invariant mass square.
In particular, ${\vec k}^2$ is given by ${\vec k}^2 = {\bf k}^2_\perp +k^2_z$ where $k_z = (x-1/2)M_0$ and 
the normalization of $\phi_R$ is given by
\be\label{eq4}
\int^1_0 dx \int\frac{d^2{\bf k}_\perp}{16\pi^3}
|\phi_R(x,{\bf k}_{\perp})|^2=P_{Q\bar{Q}}.
\ee

The covariant form of the spin-orbit wave function for the pion ($J^{PC}=0^{-+}$) is given by
\be\label{eq5} 
\chi_{\lam_1\lam_2} (x,{\bf k}_{\perp})= {\cal N} {\bar u}_{\lam_1}(k_1)\Gamma_\pi \upsilon_{\lam_2}(k_2),
\ee
where  
\be\label{eq6}
\Gamma_\pi=(A _\pi + B_\pi {/\!\!\!\!\! P})\gamma_5,
\ee
and ${\cal N}$ is the normalization constant satisfying the unitary condition
$\langle \chi_{\lam_1\lam_2}|\chi_{\lam_1 \lam_2}\rangle=1$.

Here, we set $A_\pi =M_\pi$ and $B_\pi$ being a free parameter. Explicitly, we obtain the normalized form 
of $\chi_{\lam_1\lam_2}$ as 
{
\begin{widetext}
\be\label{eq7}
\chi_{\lam_1\lam_2} (x,{\bf k}_{\perp})
=\frac{1}{\sqrt{2}\sqrt{ {\cal M}^2 \;{\bf k}^2_\perp + [m{\cal M}+  x (1-x) B_\pi \epsilon_B]^2}}\left(
\begin{array}{cc}
        -k^L {\cal M} & m{\cal M}+ x (1-x) B_\pi \epsilon_B \\ 
       -m{\cal M} - x (1-x) B_\pi \epsilon_B  & -k^R {\cal M}
      \end{array}
    \right),\;
\ee
\end{widetext}
}
\noindent
where ${\cal M}= M_\pi +2 B_\pi m$,  
$k^{R(L)}=k^x \pm i k^y$, 
and $\epsilon_B=M_\pi^2 - M^2_0$ 
corresponds to the binding energy.
The detailed derivation of Eq.~(\ref{eq7}) is given in the Appendix.
Furthermore, in the chiral limit  (i.e. $M_\pi,m\to 0$),
Eq.~(\ref{eq7}) reduces to
\be\label{eq8}
\chi^{\rm chiral}_{\lambda_1\lambda_2}
= \lim_{M_\pi,m\to 0} \chi_{\lam_1\lam_2}
=\frac{1}{\sqrt{2}}
\begin{pmatrix}
0 &  1 \\ \ $-1$ & 0
\end{pmatrix}
{\rm sgn}(-B_\pi),
\ee 
where ${\rm sgn}(-B_\pi) = -{\rm sgn}(B_\pi)$ is the sign function of $B_\pi$, i.e ${\rm sgn}(B_\pi) = 1$ for 
$B_\pi>0$, ${\rm sgn}(B_\pi) = -1$ for $B_\pi<0$ and ${\rm sgn}(0)=0$, respectively.
This reveals already the nontriviality of the axial-vector coupling, i.e. $B_\pi \neq 0$, in the chiral limit.  
We shall illustrate the way of determining the value of $B_\pi$ phenomenologically in Sec.~\ref{sec:IV}.

The operator $\Gamma_\pi$ given by Eq.~(\ref{eq6}) can be compared with those used in
the previous LFQMs using two popular schemes, i.e. the spin-averaged meson mass scheme
and the invariant meson mass scheme.
For the spin-averaged meson mass scheme used in~\cite{Dziem, CJ90}, 
$\Gamma_\pi=(M_{\rm av} + {/\!\!\!\!\! P})\gamma_5$ was taken, i.e.   the
spin-averaged meson mass  $M_{\rm av}=(\frac{1}{4}M_\pi +\frac{3}{4}M_\rho)_{\rm Exp}$
was used instead of the physical pion mass 
as mentioned earlier in Sec.~\ref{sec:I}.
For the invariant mass scheme used in~\cite{Jaus91,Choi:1997iq,Choi:1999nu,cheng04,Choi:2007yu,Choi:2007se,Choi:2017zxn},
the meson mass was mocked by the invariant mass $M_0$ and 
 $\Gamma_\pi = A_\pi \gamma_5$ was taken to yield the normalized spin-orbit wave function 
$\chi^{(M_0)}_{\lam_1\lam_2}(x,{\bf k}_{\perp})= {\cal N} {\bar u}_{\lam_1}(k_1)\gamma_5 \upsilon_{\lam_2}(k_2)$
as also mentioned in Sec.~\ref{sec:I}.
Its explicit normalized form for the pion is then given by
\be\label{eq9}
\chi^{(M_0)}_{\lam_1\lam_2}
=\frac{1}{\sqrt{2}\sqrt{{\bf k}^2_\perp+m^2}}
\begin{pmatrix}
-k^L &  m \\ -m & -k^R
\end{pmatrix}.
\ee
However, we note that the more general spin-orbit wave function given by Eq.~(\ref{eq7})
yields indeed Eq.~(\ref{eq9}) regardless of the value of $B_\pi$ in the limit $M_\pi\to M_0$ (or $\epsilon_B\to 0$)
taken in the invariant meson mass scheme. This indicates that the order of the two limits, 
i.e. the zero-binding limit ($M_\pi\to M_0$ or $\epsilon_B\to 0$)
vs. the chiral limit ($M_\pi, m\to 0$), do not commute in general regardless of the value of $B_\pi$.
While the LFQM adopting Eq.~(\ref{eq9}) has proven to be very effective in predicting various physical observables, 
its non-commutability with the chiral limit hinders the description of the chiral anomaly which determines $F_{\pi\gamma}(0)$. 
Unlike Eq.~(\ref{eq8}), Eq.~(\ref{eq9}) yields the ordinary helicity components behaving 
as $\chi^{(M_0)}_{\uparrow\downarrow}=\chi^{(M_0)}_{\downarrow\uparrow}\to 0$ 
in the chiral limit (i.e. $M_\pi, m\to 0$).

In contrast to the previous works, we now take the more general spin-orbit structure of the pion given by Eq.~(\ref{eq6}) 
which leads to the spin-orbit wave function given by Eq.~(\ref{eq7}) that is versatile enough to explore the chiral limit 
as well as the previous LFQM adopting the spin-averaged meson mass scheme or the invariant mass scheme.
\section{Application: Pion decay constant, transition and elastic form factors}
\label{sec:III}
The charged pion decay constant  is given in terms of
the matrix element of the weak current between a physical pion and the vacuum state

\be\label{eq10}
\la 0|{\bar q}\gamma^\mu(1-\gamma_5) q|\pi(P)\ra
= if_{\pi} P^\mu.
\ee
The experimental value of the pion decay constant is $f^{\rm Exp}_\pi = 130.2(1.7)$ MeV~\cite{PDG2018}.
Using the plus component ($\mu=+$) of the current, we obtain the decay constant in terms of the
valence pion LFWF~\cite{Lepage:1980fj}
\be\label{eq11}
f_\pi = 2\sqrt{2N_c}\int^1_0\;dx \int \frac{d^2{\bf k}_\perp}{16\pi^3}
\psi_\pi(x,{\bf k}_\perp),
\ee
where $N_c$ is the number of color and 
\be\label{eq12}
\psi_\pi(x,{\bf k}_\perp)
=\frac{1}{\sqrt{2}}(\chi_{\uparrow\downarrow} - \chi_{\downarrow\uparrow})\phi _{\rm R}(x,{\bf k}_\perp),
\ee
corresponds to the valence $|Q{\bar Q}\ra$ state having $J^z =S^z=L^z=0$
together with the helicity components of the spin-orbit wave function given in Eq.~(\ref{eq7}).
The twist-2 pion DA $\phi_{\pi}(x)$ results from the ${\bf k}_\perp$ integral of 
$\psi_\pi(x,{\bf k}_\perp)$ in the LF gauge $A^+=0$~\cite{Lepage:1980fj}
\be\label{eq13}
\phi_{\pi} (x) = \int^{Q^2} \frac{d^2{\bf k}_\perp}{16\pi^3}
\psi_\pi(x,{\bf k}_\perp),
\ee
and satisfies the normalization condition
\be\label{eq14}
\int^1_0 \;dx\; \phi_{\pi} (x) =\frac{f_\pi}{2\sqrt{2N_c}}.
\ee

In the chiral limit (i.e. $M_\pi,m\to 0$),  the spin-orbit part in Eq.~(\ref{eq12}) becomes 
$(\chi_{\uparrow\downarrow} - \chi_{\downarrow\uparrow})/\sqrt{2}={\rm sgn}(-B_\pi)$ (see
Eq.~(\ref{eq8})). By taking ${\rm sgn}(-B_\pi)=1$(or $B_\pi<0$),  we then obtain the decay constant and the pion DA analytically 
as
\be\label{eq15}
 f^{\rm chiral}_{\pi} = \sqrt{P_{Q\bar{Q}}} \frac{ \sqrt{3} \beta}{2^{3/4} \pi^{1/4}}\Gamma(\frac{5}{4}),
\ee
and
\be\label{eq16}
\phi^{\rm chiral}_{\pi} (x) =\frac{2\sqrt{2}  f^{\rm chiral}_\pi}{\sqrt{3} \pi} \sqrt{x(1-x)},
\ee
respectively. We should note that  this derivation of the chiral limit result could not be made 
in the case of $\Gamma_\pi =A_\pi\gamma_5$ (see Eq.~(\ref{eq9})) due to the lack of 
the axial-vector coupling.
Our chiral limit result for twist-2 pion DA given by Eq.~(\ref{eq16}) is exactly the same as the 
AdS/CFT prediction of the asymptotic DA~\cite{AdS1,AdS2,AdS3}.
We also find the exactly the same  ratio $\phi^{\rm chiral}_{\pi} (x)/ f^{\rm chiral}_\pi$
even if we use the power-law type radial wave function, 
$\phi^{\rm PL}_{R}(x,{\bf k}_{\perp}) \propto \sqrt{\partial k_z/\partial x}(1 + {\vec k}^2/\beta^2)^{-2}$
instead of using the HO wave function. This appears to indicate that the ratio $\phi^{\rm chiral}_{\pi} (x)/ f^{\rm chiral}_\pi$
is model independent.

The neutral pion transition form factor (TFF) $F_{\pi\gamma}$ for 
the $\pi^0\to\gamma^*\gamma$
transition is defined from
the matrix element of electromagnetic current $\Gamma^\mu=\la\gamma(P-q)|J^\mu_{\rm em}| \pi^0(P)\ra$
as follows:
\be\label{eq17}
\la\gamma(P-q)|J^\mu_{\rm em}|\pi^0(P)\ra = i e^2 F_{\pi\gamma}(Q^2)\ep^{\mu\nu\rho\sigma}P_\nu\vep_\rho q_\sigma,
\ee
where $P^\mu$ and $q^\mu$ are the four-momenta of the incident 
pion and virtual photon,
respectively, and $\vep_\rho$ is the transverse polarization four-vector of the final (on-shell)
photon. 

\begin{figure*}\label{fig1}
\includegraphics[height=3.5cm, width=13cm]{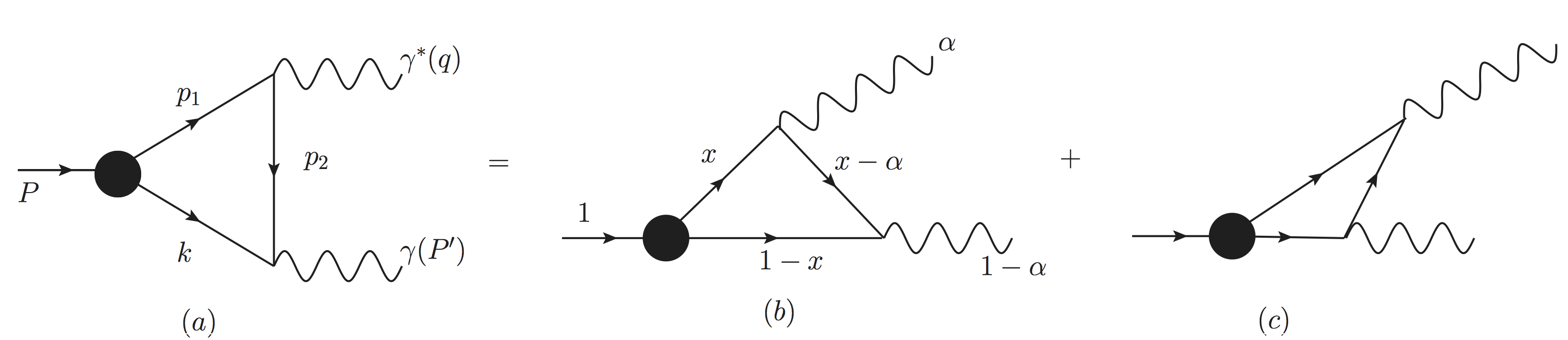}
\caption{
One-loop Feynman diagrams that contribute to $\pi^0(P)\to\gamma^*(q)\gamma(P')$. The single covariant Feynman diagram (a) is in principle the same
as the sum of the two LF time-ordered diagrams (b) and (c), respectively. 
}
\end{figure*}

As we discussed in~\cite{Choi:2017zxn}, this process is illustrated by the Feynman diagram in Fig.~1 (a), where
the intermediate quark and antiquark propagators of mass $m=m_Q=m_{\bar Q}$ carry the internal
four-momenta $p_1=P-k$, $p_2=P-q-k$, and $k$, respectively.
It is well known that the single covariant Feynman diagram Fig. 1 (a) is in general equal to the sum of the two LF time-ordered
diagrams Figs.~1 (b) and~1(c) if the $q^+\neq 0$ frame is taken. However, if the $q^+=0$ frame (but ${\bf q_\perp}\neq 0$ so 
that $q^2=q^+q^- -{\bf q}^2_\perp=-{\bf q}^2_\perp=-Q^2$) is chosen, the LF diagram 1(c) does not contribute but
only the diagram 1(b) gives exactly the same result as the covariant diagram 1(a).
However, as we found in~\cite{Choi:2017zxn},  if one takes the  $q^+ =P^+$ (or $\al=q^+/P^+=1$)  frame but with ${\bf q}_\perp=0$, 
Fig. 1(b) does not contribute but only Fig. 1(c) contributes to the total transition amplitude and shows exactly the same as the one obtained
from the $q^+=0$ frame.
While the TFF obtained from the $q^+=0$ frame is defined in the spacelike region ($q^2=-{\bf q}^2_\perp=-Q^2<0$),
the TFF obtained from  the $q^+=P^+$ frame with ${\bf q}_\perp=0$ is directly defined in the timelike region ($q^2=q^+q^->0$).
Thus,  
one can analyze the TFF in the timelike region using the $q^+=P^+$ but ${\bf q}_\perp=0$ frame 
without resorting to the analytic continuation from
spacelike region to timelike region as was did in the $q^+=0$ frame.

The  explicit form of the pion TFF obtained from the  $q^+=P^+$ frame is given by~\cite{Choi:2017zxn}  
\be\label{eq18}
F_{\pi\gamma}(q^2) = \frac{e^2_u - e^2_d}{\sqrt{2}} \frac{\sqrt{2 N_c}}{4\pi^3}\int^{1}_0
 \frac{dx}{(1-x)} \int d^2{\bf k}_\perp
 \frac{\psi_\pi(x,{\bf k}_\perp)}{M^2_0 - q^2},
 \ee
where $\psi_\pi$ is the same as  Eq.~(\ref{eq12}). 
The salient feature of Eq.~(\ref{eq18}) is that the external virtual momentum is completely decoupled from
the internal momenta ($x,{\bf k}_\perp$) and facilitates
the analysis of the timelike region $(q^2=-Q^2>0)$ due to the simple and clean pole structure,
$(M^2_0 - q^2)^{-1}$ as shown in Eq.~(\ref{eq18}). The TFF in the spacelike region can also be easily obtained by replacing $q^2$ with $-Q^2$ 
in $(M^2_0 - q^2)^{-1}$ and was shown to be exactly the same as the result obtained from the  $q^+=0$ frame~\cite{Choi:2017zxn}.
We note that the leading order QCD result~\cite{Lepage:1980fj} for $F_{\pi\gamma}(Q^2)$ with $\phi_\pi(x)=6x(1-x)$, 
so called Brodsky-Lepage (BL) limit
at the asymptotic region (i.e. $Q^2\to\infty$), is given by $Q^2 F_{\pi\gamma}(Q^2)=\sqrt{2}f_\pi \simeq 0.185$ GeV. 
As one can see clearly from Eq.~(\ref{eq18}), our model satisfies the scaling behavior $Q^2 F_{\pi\gamma}(Q^2)\to$ constant
as $Q^2\to\infty$. But how large $Q^2$ should be to reach the scaling behavior is related with the model parameters as we shall
discuss in Sec.~\ref{sec:IV}.
On the other hand, the decay width for $\pi^0\to\gamma\gamma$ is obtained from the TFF at $Q^2=0$ via
\be\label{eq19}
\Gamma_{\pi^0\to\gamma\gamma}=\frac{\pi}{4}\alpha^2_{\rm em} M^3_\pi |F_{\pi\gamma}(0)|^2,
\ee
where $\al_{\rm em}$ is the fine structure constant. The form factor $F_{\pi\gamma}(0)$ is also well described by
the following ABJ anomaly (or the chiral anomaly)~\cite{ABJ}
\be\label{eq20}
F^{\rm ABJ}_{\pi\gamma}(0)=\frac{1}{2\sqrt{2}\pi^2 f_\pi},
\ee
which results in $F^{\rm ABJ}_{\pi\gamma}(0)\simeq 0.276$ GeV$^{-1}$ for $f^{\rm Exp}_\pi\simeq 130$ MeV
agreeing with
the experimental data $F^{\rm Exp}_{\pi\gamma}(0)= 0.272(3)$ GeV$^{-1}$ within a few percent. 

From Eq.~(\ref{eq18}), we obtain the TFF in the exact chiral limit ($M_\pi,m\to 0$) and its
 analytic form is given by
\be\label{eq21}
F^{\rm chiral}_{\pi\gamma}(Q^2)
= \sqrt{P_{Q\bar{Q}}}
\frac{\Gamma \left(\frac{5}{4}\right) e^{\frac{Q^2}{8 \beta ^2}} \sqrt{\frac{Q}{\beta ^3}} \Gamma \left(-\frac{1}{4},\frac{Q^2}{8 \beta ^2}\right)}{4 \sqrt{3} \sqrt[4]{\pi }}.
\ee
In particular, the TFF at $Q^2=0$ is obtained as
\be\label{eq22}
F^{\rm chiral}_{\pi\gamma}(0)= \sqrt{P_{Q\bar{Q}}} \frac{\Gamma(\frac{5}{4})}{2\sqrt{3}(2\pi)^{1/4}\beta} 
= \frac{ \sqrt{ \frac{\pi^3}{32}} [\Gamma(\frac{1}{4})]^2 P_{Q{\bar Q}}}{2\sqrt{2}\pi^2 f^{\rm chiral}_\pi},
\ee
where we used Eq.~(\ref{eq15}) to obtain the second expression of Eq.~(\ref{eq22}).
Equating Eq.~(\ref{eq22}) with Eq.~(\ref{eq20}),  we find that $P_{Q{\bar Q}}<0.1$ in the chiral limit of
our model is required to fit both $f^{\rm Exp}_\pi$ and  $\Gamma^{\rm Exp}_{\pi^0\to\gamma\gamma}$
correctly. This indicates a significant higher Fock-state contribution in the chiral limit.
This point has been also discussed in the LF holographic QCD based on the AdS/CFT correspondence
in which $P_{Q{\bar Q}}=0.5$ was estimated to describe simultaneously
$\Gamma_{\pi^0\to\gamma\gamma}$ and the pion TFF at the asymptotic limit.
As we shall show in Sec.~\ref{sec:IV}, the probability $P_{Q{\bar Q}}$ increases as the
quark mass increases indicating the saturation of the LF Fock-state expansion 
with the lower Fock-state contribution as the  so-called ``current" quarks get amalgamated with themselves 
to form the constituent quark degrees of freedom. In our numerical calculation of Sec. IV, 
we shall analyze the mass variation effect as $Q^2$ gets evolved
and also compare with the result~\cite{AdS4} 
obtained from the LF holographic QCD based on the AdS/CFT correspondence.

The electromagnetic form factor  $F_\pi(Q^2)$ of a pion is defined by the matrix elements of the
current operator $J^\mu_{\rm em}$:

\be\label{eq23}
\la P'|J^\mu_{\rm em}|P\ra = (P+P')^\mu F_\pi(Q^2),
\ee
where 
$q=P-P'$ is the four momentum transfer.

Our calculation for $F_\pi(Q^2)$ is carried out using the standard LF frame 
($q^{+}=0$).
The charge form factor of the pion can then be expressed as the convolution of the initial and final state LF wave functions
for the ``$+$" component of the current operator $J^{\mu}_{\rm em}$ as follows
{\begin{widetext}
\bea\label{eq24}
F_\pi(Q^{2})
&=&\int^{1}_{0}dx\int \frac{d^{2}{\bf k}_{\perp}}{16\pi^3}
\Psi^*_\pi(x,{\bf k}'_\perp)\Psi_\pi(x,{\bf k}_\perp)
\nonumber\\
&=& \int^{1}_{0}dx\int \frac{d^{2}{\bf k}_{\perp}}{16\pi^3}
\phi^{*}_R(x,{\bf k}'_{\perp}) \phi_R(x,{\bf k}_{\perp})
\frac{{\cal M}^2 {\bf k}_{\perp}\cdot{\bf k}'_{\perp} + [m {\cal M} + x(1-x) B_\pi\epsilon_B][m {\cal M} + x(1-x) B_\pi\epsilon'_B] }
{ \sqrt{{\cal M}^2 {\bf k}^{2}_{\perp} +  [m {\cal M} + x(1-x) B_\pi\epsilon_B]^2}
\sqrt{{\cal M}^2 {\bf k}'^{2}_{\perp} +  [m {\cal M} + x(1-x) B_\pi\epsilon'_B]^2} },
\eea
\end{widetext}
}
\noindent
where ${\bf k}'_{\perp}={\bf k}_{\perp} + (1-x){\bf q}_{\perp}$ and 
$\epsilon'_B$ is the same as $\epsilon_B$ but with the replacement of ${\bf k}_{\perp}$ with ${\bf k}'_{\perp}$.
One can also easily find that the spin-orbit term in Eq.~(\ref{eq24}) becomes
1 in the chiral limit (i.e. $M_\pi, m\to 0$).
The charge radius of the pion can be calculated by
$\la r^{2}_\pi\ra$=$-6dF_\pi(Q^{2})/dQ^{2}|_{Q^{2}=0}$.

\section{Numerical Results}
\label{sec:IV}

\begin{figure}
\includegraphics[width=6.5cm, height=5cm]{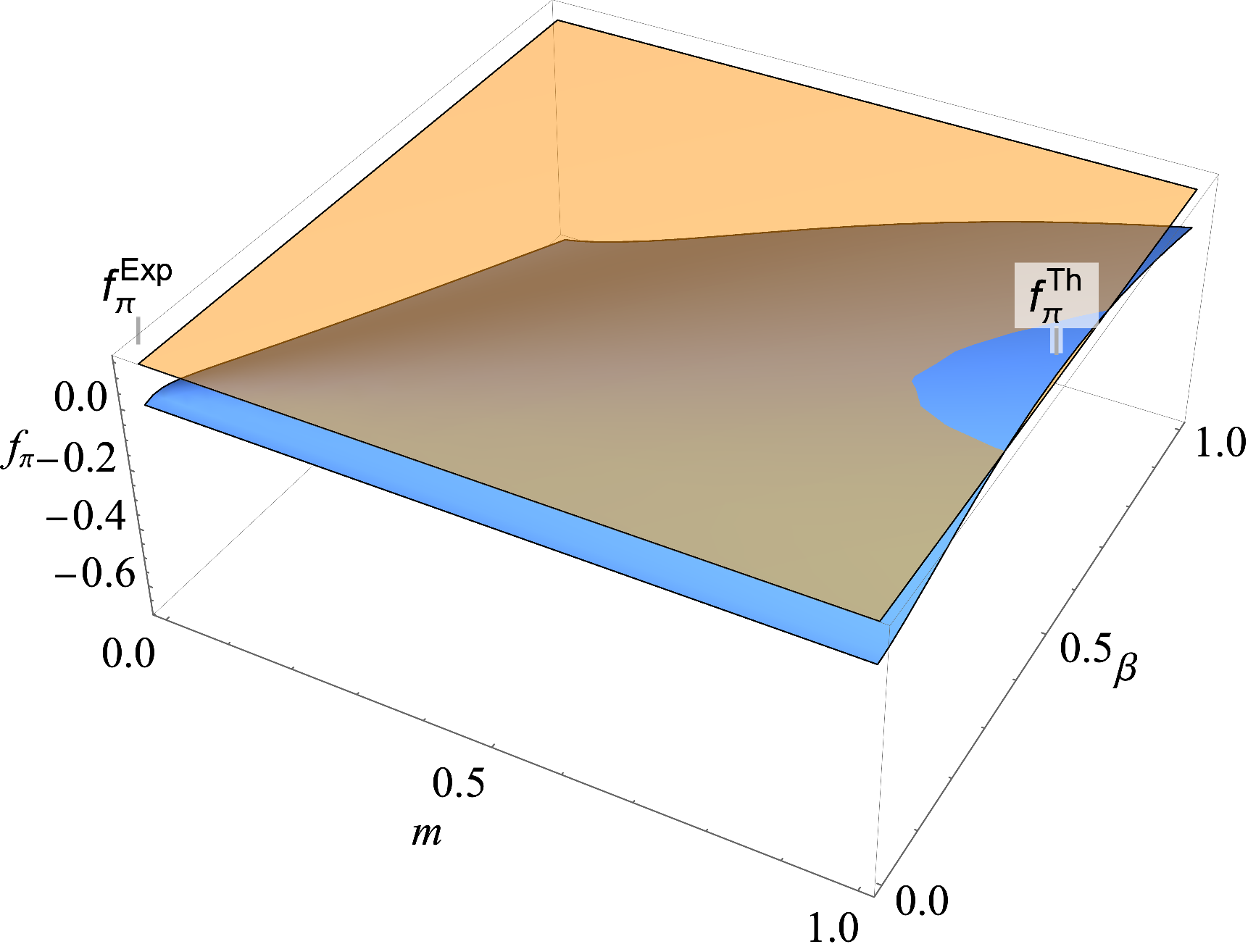}
\includegraphics[width=6.5cm, height=5.5cm]{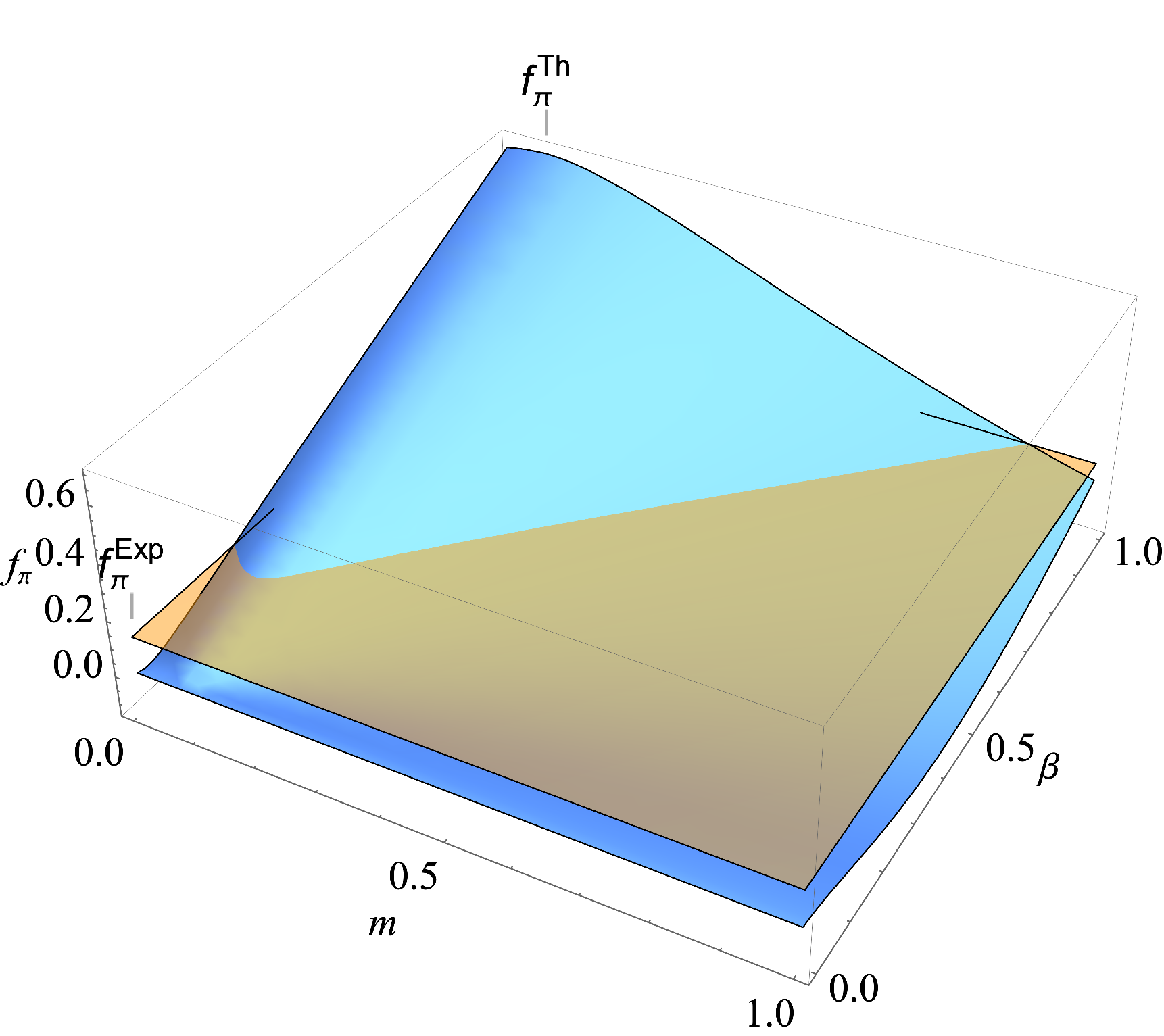}
\caption{ \label{fig2} Possible solution sets for $(m, \beta)$ satisfying $f^{\rm Th}_\pi = f^{\rm Exp}_\pi$ obtained from
the pion vertex $\Gamma_\pi=(M_\pi + B_\pi/\!\!\!\!\!P)\gamma_5$  with the $B_\pi=1$ (upper panel) and $B_\pi=-1$ (lower panel).}
\end{figure}

\begin{figure}
\begin{center}
\includegraphics[width=7cm, height=7cm]{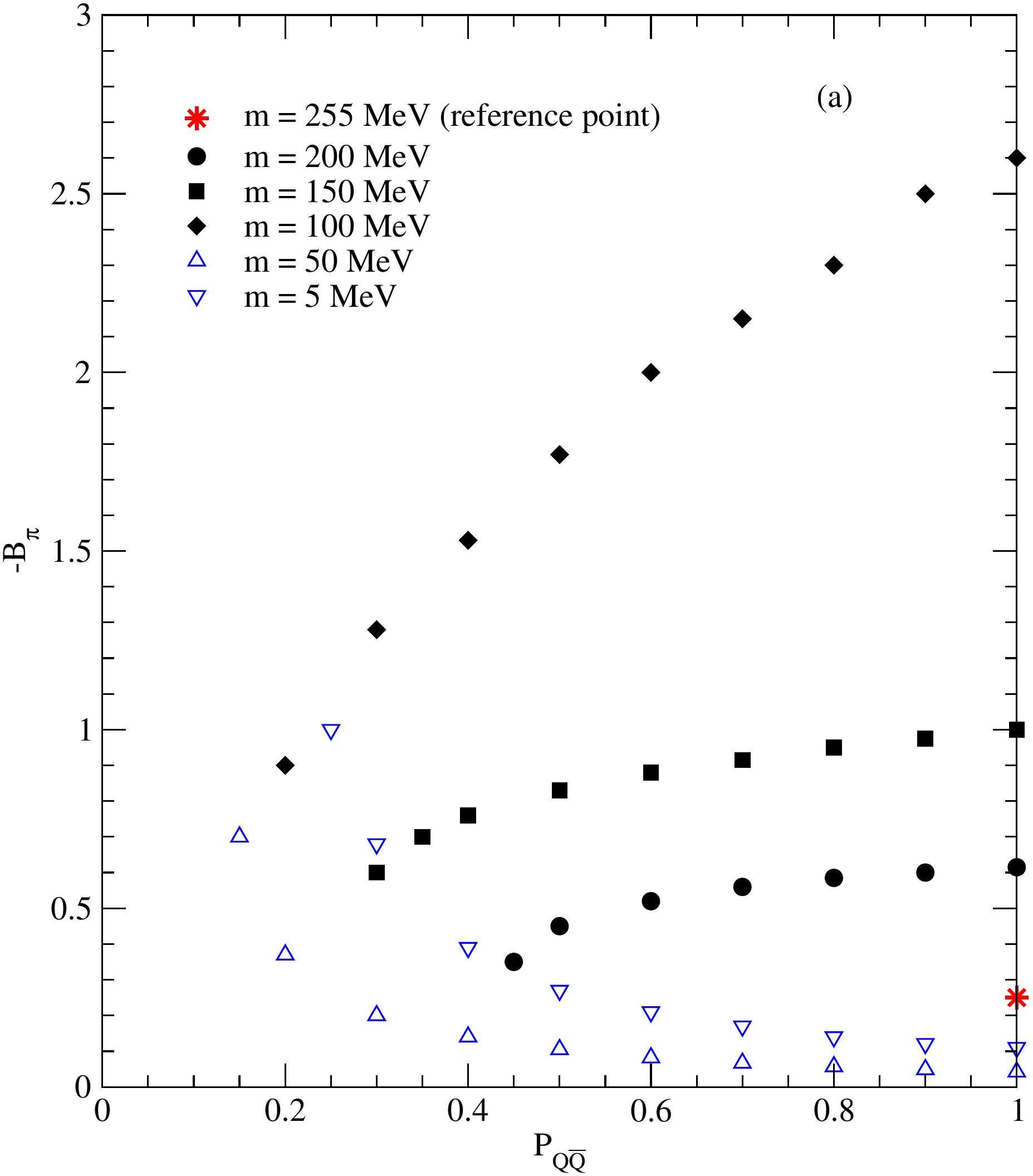}
\\
\includegraphics[width=7cm, height=7cm]{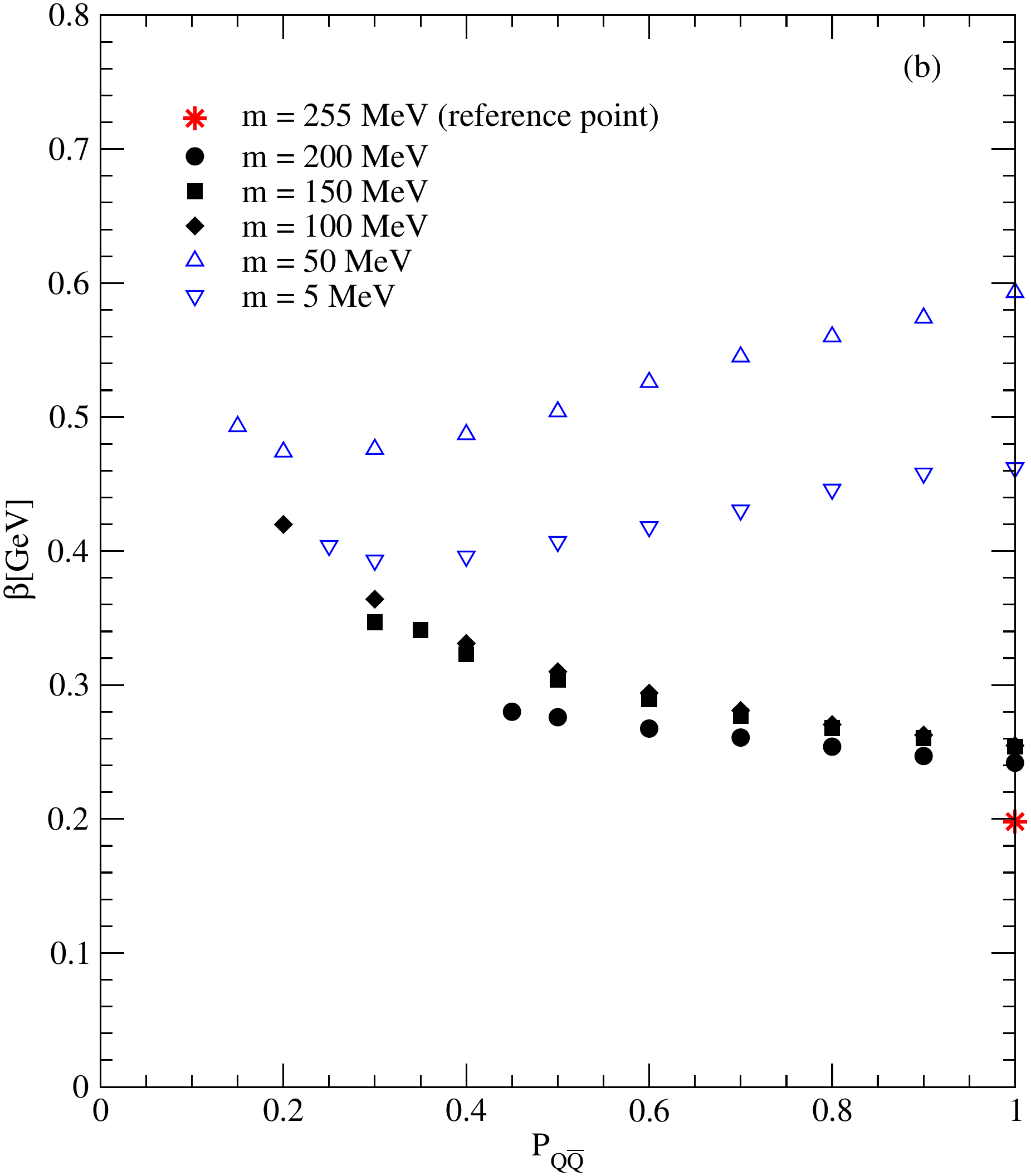}
\caption{\label{fig3} Possible solution sets for ($-B_\pi$ vs. $P_{Q{\bar Q}}$) (a)
and ($\beta$ vs. $P_{Q{\bar Q}}$) (b) for given quark mass $m=(255,200,150,100,50,5)$ MeV and $M_\pi=135$ MeV
satisfying $f^{\rm Exp}_\pi$ and $F^{\rm Exp}_{\pi\gamma}(0)$, simultaneously. We set the solution
of $m=255$ MeV with $P_{Q\bar{Q}}=1$ as a reference point.
}
\end{center}
\end{figure}

 Laying out all the formulae for our model description and its application to $f_\pi$, $F_{\pi \gamma}(Q^2)$ and $F_\pi(Q^2)$ in Sec.~\ref{sec:II} and Sec.~\ref{sec:III}, respectively, we have already discussed the critical role of chiral anomaly in constraining the model parameters. In particular, we noticed 
not only the nontriviality of the axial-vector coupling, i.e. $B_\pi \neq 0$, in the chiral limit  
but also the negativity of the axial-vector coupling, i.e. $B_\pi < 0$, to dictate 
the model independence of the ratio $\phi^{\rm chiral}_{\pi} (x)/ f^{\rm chiral}_\pi$
and the consistency with the AdS/CFT prediction of asymptotic DA.
Moreover, the probability of the lowest LF Fock state $P_{Q{\bar Q}}$ should diminish 
in the chiral limit to obtain both the chiral anomaly (i.e. $\Gamma^{\rm Exp} _{\pi^0\to\gamma\gamma}$)
and $f^{\rm Exp}_\pi$ correctly, indicating a significant higher Fock-state contribution in the chiral limit.
In this section, we present our numerical results and discuss the consistency of  
the constraints on the model parameters with the chiral anomaly that we discussed in previous sections. 

To find the optimum model parameters $(m,\beta, B_\pi)$, we first take $P_{Q{\bar Q}}=1$ and fit the two experimental  data, (1) pion decay constant $f^{\rm Exp}_\pi=130.2(1.7)$ MeV, and (2) $F^{\rm Exp}_{\pi\gamma}(0) = 0.272 (3)$ GeV$^{-1}$, simultaneously.
For an illustration, we show in Fig.~\ref{fig2}  the possible solution sets for $(m,\beta)$ satisfying $f^{\rm Th}_\pi = f^{\rm Exp}_\pi$ 
when $B_\pi =+1$ (upper panel) and $-1$ (lower panel) for given pion physical mass $M_\pi=135$ MeV.
As one can see from Fig.~\ref{fig2},
while the negative sign of $B_\pi$ has the solution set (i.e. overlap line between the blue and peach colors) covering all the
possible range of $0\leq (m,\beta)\leq 1$ GeV, the positive sign of $B_\pi$ has the solution set covering severely restricted range with rather 
unusually large $u(d)$-quark mass (i.e. $m\geq 0.7$ GeV). 
The restriction on the model parameters for the case of $B_\pi =+1$ appears in line with the unusually large Mock meson 
mass $M_{\rm av}=(\frac{1}{4}M_\pi +\frac{3}{4}M_\rho)_{\rm Exp}\approx 612$ MeV in the spin-averaged mass scheme~\cite{Dziem,CJ90} for the consistency with the experimental data. 
As already indicated in the results of $f^{\rm chiral}_\pi$ and $\phi^{\rm chiral}_{\pi} (x)$ in the chiral limit given by Eqs.~(\ref{eq15}) and (\ref{eq16}), the negativity of $B_\pi$, i.e. $B_\pi < 0$, is essential for the consistency in the chiral limit. Varying the value of $P_{Q{\bar Q}}$, i.e. $1>P_{Q{\bar Q}}>0$, we have confirmed that the value of $B_\pi$ should be taken to be negative 
in order to make a link to the chiral limit. 

In Fig.~\ref{fig3}, we show the possible solution sets for the model parameters depending on $(m, \beta, B_\pi, P_{Q{\bar Q}})$ 
 for given quark mass $m=(255,200,150,100,50,5)$ MeV and $M_\pi=135$ MeV, i.e. 
($-B_\pi$ vs. $P_{Q{\bar Q}}$) in Fig.~\ref{fig3}(a)
and ($\beta$ vs. $P_{Q{\bar Q}}$) in Fig.~\ref{fig3}(b),
which were obtained by fitting both $f^{\rm Exp}_\pi=130.2(1.7)$ MeV
and $F^{\rm Exp}_{\pi\gamma}(0)=0.272(3)$ GeV$^{-1}$ simultaneously. 
In our previous work~\cite{Choi:2017zxn} with $B_\pi = 0$, i.e. $\Gamma_\pi = A_\pi \gamma_5$, 
the quark mass $m=220$ MeV and the Gaussian parameter $\beta=0.3659$ GeV
were taken from our earlier LFQM~\cite{Choi:1997iq} spectroscopic analysis of the ground state pseudoscalar and vector meson nonets based on the variational principle. In the scope of present work involving only the pion, however, we do not attempt a spectroscopic analysis but focus on the effect of nonzero axial vector coupling ($B_\pi < 0$) for the consistency with the chiral anomaly. 
For this purpose, we first set our reference parameter set with  $P_{Q\bar{Q}}=1$ and $B_\pi = -0.25$ 
which is a rather small axial vector coupling compared to the pseudoscalar coupling
and find the corresponding optimum values of $m$ and $\beta$ to fit both $f^{\rm Exp}_\pi$
and $F^{\rm Exp}_{\pi\gamma}(0)$. Then, by reducing the quark mass $m$ from this reference point and again fitting both $f^{\rm Exp}_\pi$ and  $F^{\rm Exp}_{\pi\gamma}(0)$ simultaneously, 
we obtain the rest of parameter sets shown in Fig.~\ref{fig3}. We mark the reference parameter set 
by asterisk ({\color{red}$\ast$}) in Fig.~\ref{fig3}, i.e. $(M_\pi, m, \beta)=(0.135, 0.255, 0.1980)$ GeV and
$(B_\pi, P_{Q{\bar Q}})=(-0.25,1)$, with which we get  
$F_{\pi\gamma}(0)=P_{Q{\bar Q}}/(2\sqrt{2}\pi^2 f_\pi)=0.271$ GeV$^{-1}$ and $f_\pi=130.4$ MeV
close enough to $F^{\rm Exp}_{\pi\gamma}(0)=0.272(3)$ GeV$^{-1}$ and $f^{\rm Exp}_\pi=130.2(1.7)$ MeV for our purpose in this work. In comparison with the value $\beta=0.3659$ GeV in the absence of the axial vector coupling $B_\pi = 0$~\cite{Choi:2017zxn}, the value $\beta=0.1980$ GeV in the reference parameter set is somewhat reduced with the contribution of axial vector coupling $B_\pi = -0.25$, while the quark mass $m=255$ MeV still represents the ordinary constituent quark picture
in our reference point ``{\color{red}$\ast$}". In reducing the quark mass $m$ from this reference point to fit
both $f^{\rm Exp}_\pi$ and  $F^{\rm Exp}_{\pi\gamma}(0)$ simultaneously, we ultimately reached 
the parameter set $(M_\pi, m)=(135, 5)$ MeV reproducing the Gell-Mann-Oakes-Renner (GMOR) relation~\cite{GMOR}, i.e. $M^2_\pi f^2_\pi = -2(m_q + m_{\bar q}) \la q\bar{q}\ra$, where $\la q{\bar q}\ra =-(250$ MeV)$^3$ with the ``current" quark mass $m=m_q=m_{\bar q}$. 
For the fixed value of the pion mass, i.e. $M_\pi = 0.135$ GeV, we distinguish the two different cases of the quark-antiquark bound state, i.e. $M_\pi < 2m$ vs. $M_\pi > 2m$, 
and call them as the ``constituent quark picture" vs. the ``current quark picture", respectively.
In Fig.~\ref{fig3}, the parameter sets corresponding to $M_\pi < 2m$ and $M_\pi > 2m$ cases
are denoted by black and blue data, 
respectively. 

From the results shown in Fig.~\ref{fig3}, we summarize our main findings for the model parameters as follows:
(1) The minimum probability $P^{\rm min}_{Q{\bar Q}}$ exists for a given quark mass
satisfying both $f^{\rm Exp}_\pi$ and $F^{\rm Exp}_{\pi\gamma}(0)$ simultaneously, e.g. 
$P^{\rm min}_{Q{\bar Q}}=(0.45, 0.25)$ for $m=(200, 5)$ MeV etc.
This result is in line with the trend that the probability $P_{Q{\bar Q}}$ increases as the quark mass increases indicating the saturation of the LF Fock-state expansion 
with the lower Fock-state contribution as the current quarks get amalgamated with themselves  
to form the constituent quark degrees of freedom.
(2) For the quark masses satisfying $M_\pi < 2 m$ (i.e. constituent quark picture), 
the Guassian parameter $\beta$ gets larger as $P_{Q{\bar Q}}$ decreases. 
This indicates that the spatial size of the lowest Fock state gets smaller as the higher Fock states
contribute more. For a given quark mass $m$, the axial vector coupling $-B_\pi$ gets also reduced
as the higher Fock states contribute more, i.e. $P_{Q{\bar Q}}$ decreases. 
For a fixed $P_{Q{\bar Q}}$, however, we notice that $-B_\pi$ increases quite significantly as $m$ decreases while $\beta$ values do not change much indicating only marginal size reduction  in the lowest Fock state with the reduction of mass $m$.
(3) For the quark masses satisfying $M_\pi > 2 m$ (i.e. current quark picture), 
 $\beta$ values are in general greater for the current quark mass than the constituent one for given $P_{Q{\bar Q}}$ indicating that the spatial size of the lowest Fock state consisted 
of the current quark is smaller than the one consisted of the constituent quark.
As $P_{Q{\bar Q}}$ decreases, however, $\beta$ values get reduced down to those in the constituent quark picture indicating that the spatial size of the lowest Fock state consisted 
of the current quark gets larger as the higher Fock states contribute more.
The similar merge of the axial vector coupling $B_\pi$ between the current quark picture
and the constituent picture appears as $P_{Q{\bar Q}}$ decreases in
the upper panel Fig.~\ref{fig3}(a).  It is indeed fascinating to observe the merge 
of the parameter sets between the current quark picture and the constituent picture
as $P_{Q{\bar Q}}$ decreases both in Fig.~\ref{fig3}(a) and Fig.~\ref{fig3}(b).
It seems to indicate a nontrivial dynamic saturation process of 
the LF Fock-state expansion occurring as the current quarks get amalgamated with themselves  
to form the constituent quark degrees of freedom according to these results.

For the case of exact chiral limit ($M_\pi=m=0$), our results for any physical observables are
independent of $B_\pi$ as far as it is negative nonzero value ($B_\pi <0$) and depend only on $(\beta, P_{Q{\bar Q}})$, 
which were obtained as $\beta=0.6685$ GeV and $P_{Q{\bar Q}}=0.078$ by fitting
 both $f^{\rm Exp}_\pi$ (see Eq.~(\ref{eq15})) and $F^{\rm Exp}_{\pi\gamma}(0)$  (see Eq.~(\ref{eq22}))  simultaneously.

Table~\ref{t1} shows our  typical model parameters $(B_\pi, \beta)$ depending on the variation of $(M_\pi, m)$ and
$P_{Q{\bar Q}}$ used in the analysis of the twist-2 DA $\phi_\pi(x)$,  the transition form factor $F_{\pi\gamma}(Q^2)$, and the electromagnetic
form factor
$F_\pi(Q^2)$.
Among many possible solutions satisfying both $f^{\rm Exp}_\pi$ and $F^{\rm Exp}_{\pi\gamma}(0)$ as shown in Fig.~\ref{fig3},
we select a few parameter sets $(M_\pi, m)=\{(135,255), (135, 150), (135,50), (0,0)\}$ MeV
corresponding to the variation of the probability $P_{Q{\bar Q}}=\{ 1, 0.3, 0.15, 0.078\}$ 
in order to estimate the mass variation effect on both $F_{\pi\gamma}(Q^2)$  and $F_\pi(Q^2)$  form factors.
\begin{table}
\caption{Model parameters $(B_\pi, \beta)$ depending on the variation of $(M_\pi, m)$ and
$P_{Q{\bar Q}}$. We denote $(M_\pi, m, \beta, f_\pi)$  in unit of MeV.
}
\label{t1}
\begin{tabular}{llllll} \hline\hline
$(M_\pi,m)$ & $P_{Q\bar{Q}}$ & $B_\pi$ & $\beta$  & $f^{\rm Th}_\pi$ & $F^{\rm Th}_{\pi\gamma}(0)$   [GeV$^{-1}$]   \\
\hline
(135,255) & 1 & $-0.25$ & 198.0 &   130.4&  0.271  \\
\hline
(135,150) & $0.3$ & $-0.60$ & 346.9  &   $130.6$  &  $0.272$ \\
\hline
(135,50)  &  $0.15$  & $-0.7$ & 493.0 &   $130.7$ & $0.271$   \\
\hline
(0,0)  &  0.078  & $<0$ & 668.5 &   130.9 & 0.276   \\
\hline
Exp.~\cite{PDG2018}  & $-$ & $-$ & $-$ & 130.2(1.7)&  0.272(3)    \\
\hline\hline
\end{tabular}
\end{table}

\begin{figure}
\includegraphics[width=7cm, height=7cm]{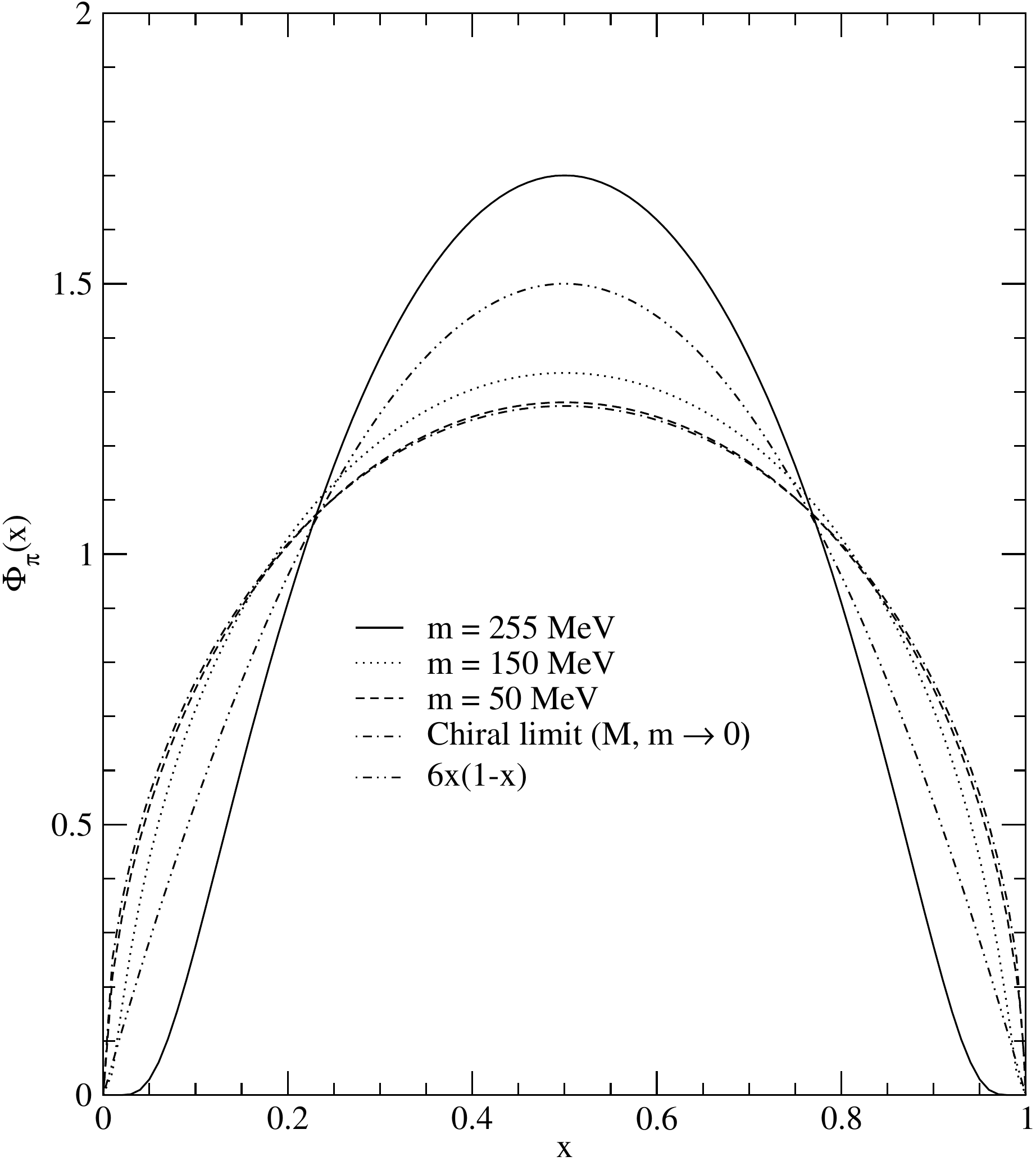}
\caption{\label{fig4}The normalized pion DA $\Phi_\pi(x)$
obtained from the sets of $(M_\pi,m)=\{(135, 255), (135, 150), (135, 50), (0,0)\}$ MeV 
with  $P_{Q{\bar Q}}=\{1, 0.3, 0.15, 0.078 \}$ compared with the asymptotic DA.}
\end{figure}

Using these typical parameter sets in Table~\ref{t1},
we first show the normalized twist-2 pion DA  $\Phi_\pi(x)$ satisfying $\int^1_0 dx\; \Phi_\pi(x)=1$ and 
compare them with the asymptotic DA, $\Phi^{\rm asy}_\pi=6x(1-x)$ in Fig.~\ref{fig4}.
The twist-2 pion DA with larger quark mass
such as $m=255$ MeV
is strongly suppressed in the vicinity of endpoints ($x=0,1$) but the DA shows broader shape  than 
the asymptotic DA (double-dot-dashed line) as the quark mass is getting smaller.
Our chiral limit result (dot-dashed line) is exactly the same as the AdS/CFT prediction 
of the asymptotic DA~\cite{AdS1,AdS2,AdS3}.
We obtain $\Phi_\pi(1/2) = (1.70, 1.34, 1.28)$ for $m=(255, 150, 50)$ MeV 
and $\Phi^{\rm chiral}_\pi(1/2) =1.27$ for $(M_\pi, m)=(0,0)$, which should be compared with  $\Phi^{\rm asy}_\pi(1/2) =1.5$ 
as well as other theoretical predictions such as 
$\Phi^{\rm SR}_\pi(1/2) =1.2\pm 0.3$  obtained from QCD sum rules~\cite{SR_BF}, 
$\Phi^{\rm RL(DB)}_\pi(1/2) =1.16(1.29)$ from Dyson-Schwinger equation approach using the
rainbow-ladder truncation (RL) and the dynamical chiral-symmetry breaking improved (DB)
kernels~\cite{DSE1, DSE2}, and $\Phi^{\rm LFQM}_\pi(1/2) =1.25$ from our
LFQM using the spin structure given by Eq.~(\ref{eq9}) and the linear
confining potential model parameters~\cite{Choi:2007yu}.
One can also define the expectation value of the longitudinal momentum, so called $\xi=2x-1$ moments 
as
\be\label{eq25}
\la \xi^n \ra= \int^1_0 dx \xi^n \Phi_\pi (x).
\ee
The odd power of $\xi$-moments for the pion DA are zero due
to the isospin symmetry, and the first nonzero moment $(n=2)$ is obtained as
$\la \xi^2 \ra = (0.155, 0.230, 0.247)$ for $m=(255, 150,50)$ MeV with $P_{Q{\bar Q}}=(1, 0.3, 0.15)$, 
and 0.250 in the exact chiral limit ($M_\pi, m\to 0$), whereas it is
0.20 for the asymptotic DA. 
Our result for $\la \xi^2\ra$ gets larger as the quark mass and the probability are getting smaller
and reaches maximum value $\la \xi^2\ra^{\rm max}=0.25$ in the chiral limit.
Especially, our result for $\la \xi^2\ra$ obtained from the current quark picture (i.e. $2m< M_\pi$)
are quite comparable with  $\la \xi^2 \ra^{\rm LFQM}=0.24$
obtained from our previous LFQM~\cite{Choi:2007yu} 
and other theoretical predictions such as
$\la \xi^2 \ra^{\rm Lat}=0.27\pm 0.04$ from Lattice QCD~\cite{Lattice06}, 
$\la \xi^2 \ra^{\rm RL(DB)}=0.28(0.25)$ from~\cite{DSE1, DSE2}.

The profiles of normalized twist-2 pion DA  $\Phi_\pi(x)$ 
shown in Fig.~\ref{fig4} exhibit a dramatic difference in the endpoint behaviors
near $x=0$ and $1$ between the two typical parameter sets 
$(M_\pi, m)=\{(135,255), (135,50)\}$ MeV which represent 
the constituent quark picture and the current quark picture, respectively.  
This difference in the end-point behaviors would be consequential 
in describing $F_{\pi\gamma}(Q^2)$ and $F_\pi(Q^2)$. 
Moreover, the current quark picture gets closer to the chiral limit
than the constituent quark picture does as indicated in  
almost indistinguishable profiles between the two parameter sets $(M_\pi, m)=\{(135,50),(0,0)\}$ MeV
shown in Fig.~\ref{fig4}. 
These results motivate us to explore the analysis of $F_{\pi\gamma}(Q^2)$ and $F_\pi(Q^2)$
for both low- and high- $Q^2$ regions, estimating the mass variation effect on 
the $Q^2$ evolution of these form factors.
Since the exact form of the quark mass evolution is still not known in the light-front dynamics,  
we prescribe the mixing of different mass eigenstates as 
 our first attempt to estimate the quark mass variation effect by taking
$\la\Psi^\pi_{m'} | \Psi^\pi_{m}\ra =\delta_{m'm} \sqrt{P_{m'} P_m}=\delta_{m'm}P_m$ with $P_m$ given by the corresponding $P_{Q\bar{Q}}$ for 
the mass $m$ in Table~\ref{t1}. 

With this idea in mind, we try to implement the quark mass variation in our LFQM to describe
$F_{\pi\gamma}(Q^2)$ and $F_\pi(Q^2)$ for both low- and high- $Q^2$ regions.
Namely, we obtain $F_{\pi\gamma}(Q^2)$ and $F_\pi(Q^2)$ by combining the contribution from 
the LF quark degrees of freedom at the reference point with $m_{\rm ref}=m=255$ MeV and $P_{Q{\bar Q}}=P_{m_{\rm ref}}=1$ 
with another contribution from the LF quark degrees of freedom with the reduced $m$ and $P_m$. 
For instance, the form factors 
$F^{(m_{\rm ref}, m)}_{\pi\gamma}(Q^2)$ and $F^{(m_{\rm ref}, m)}_\pi(Q^2)$ 
obtained from the mixing of 
the LF quark degrees of freedom with $m_{\rm ref}=255$ MeV and $P_{m_{\rm ref}}=1$ and
the LF quark degrees of freedom with $m=150$ MeV with $P_m=0.3$ in Table~\ref{t1} are 
respectively given by
\be\label{eq26a}
F^{(m_{\rm ref}, m=150)}_{\pi\gamma} (Q^2) =\frac{ \sqrt{1- {\tilde P}_m} 
F^{(m_{\rm ref})}_{\pi\gamma}+\sqrt{{\tilde P}_m} F^{(m=150)}_{\pi\gamma}}
{\sqrt{1- {\tilde P}_m} + \sqrt{ {\tilde P}_m} },
\ee
and
\be\label{eq26b}
F^{(m_{\rm ref}, m=150)}_{\pi} (Q^2) = (1- {\tilde P}_m) F^{(m_{\rm ref})}_\pi+ {\tilde P}_m F^{(m=150)}_\pi,
\ee
where the renormalized probability is denoted as ${\tilde P}_m = P_m/(P_{m_{\rm ref}}+P_m)=0.3/1.3 \approx 0.23$.
Since our prediction of the TFF satisfies $F^{\rm Th}_{\pi\gamma}(0)\simeq F^{\rm Exp}_{\pi\gamma}(0)$ for any quark mass
as shown in Table~\ref{t1},
our value of $F^{(m_{\rm ref}, m)}_{\pi\gamma}(Q^2)$
also satisfies $F^{(m_{\rm ref}, m)}_{\pi\gamma}(0)\simeq F^{\rm Exp}_{\pi\gamma}(0)$.
Also, $F^{(m)}_\pi(0)=1$ for any quark mass $m$ as given by Eq.~(\ref{eq24}) so that
the normalization $F^{(m_{\rm ref}, m)}_{\pi} (0)=1$ for the electromagnetic form factor in Eq.~(\ref{eq26b})
is always satisfied.

\begin{figure*}
\begin{center}
\includegraphics[width=7cm, height=7cm]{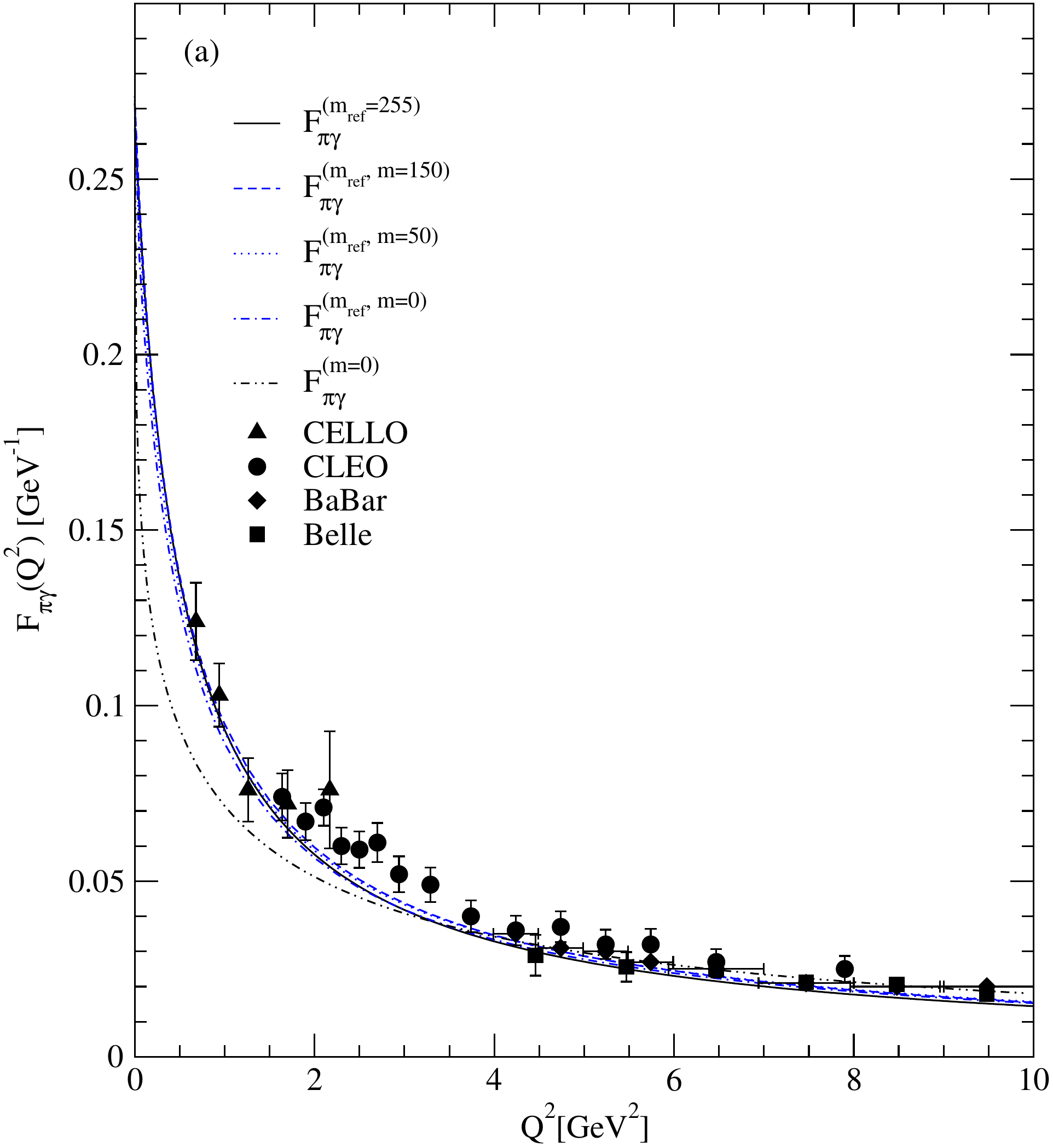}
\includegraphics[width=7cm, height=7cm]{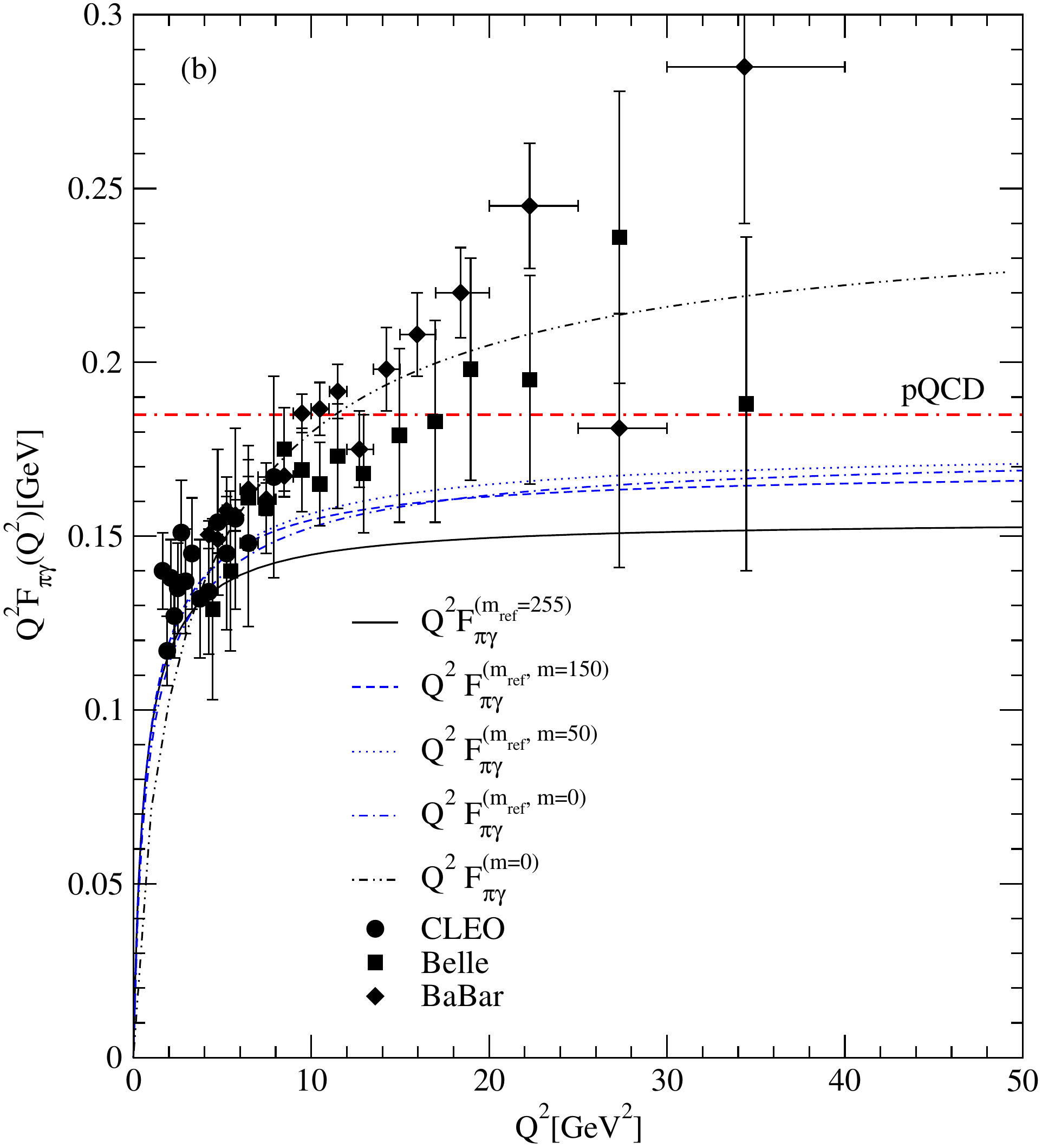}
\caption{\label{fig5} Predictions of (a) $F_{\pi\gamma}(Q^2)$ and (b) $Q^2 F_{\pi\gamma}(Q^2)$. The experimental data are taken 
from~\cite{Behrend:1990sr,Gronberg:1997fj,Uehara:2012ag,Aubert:2009mc}.
}
\end{center}
\end{figure*}

\begin{figure*}
\begin{center}
\includegraphics[width=7cm, height=7cm]{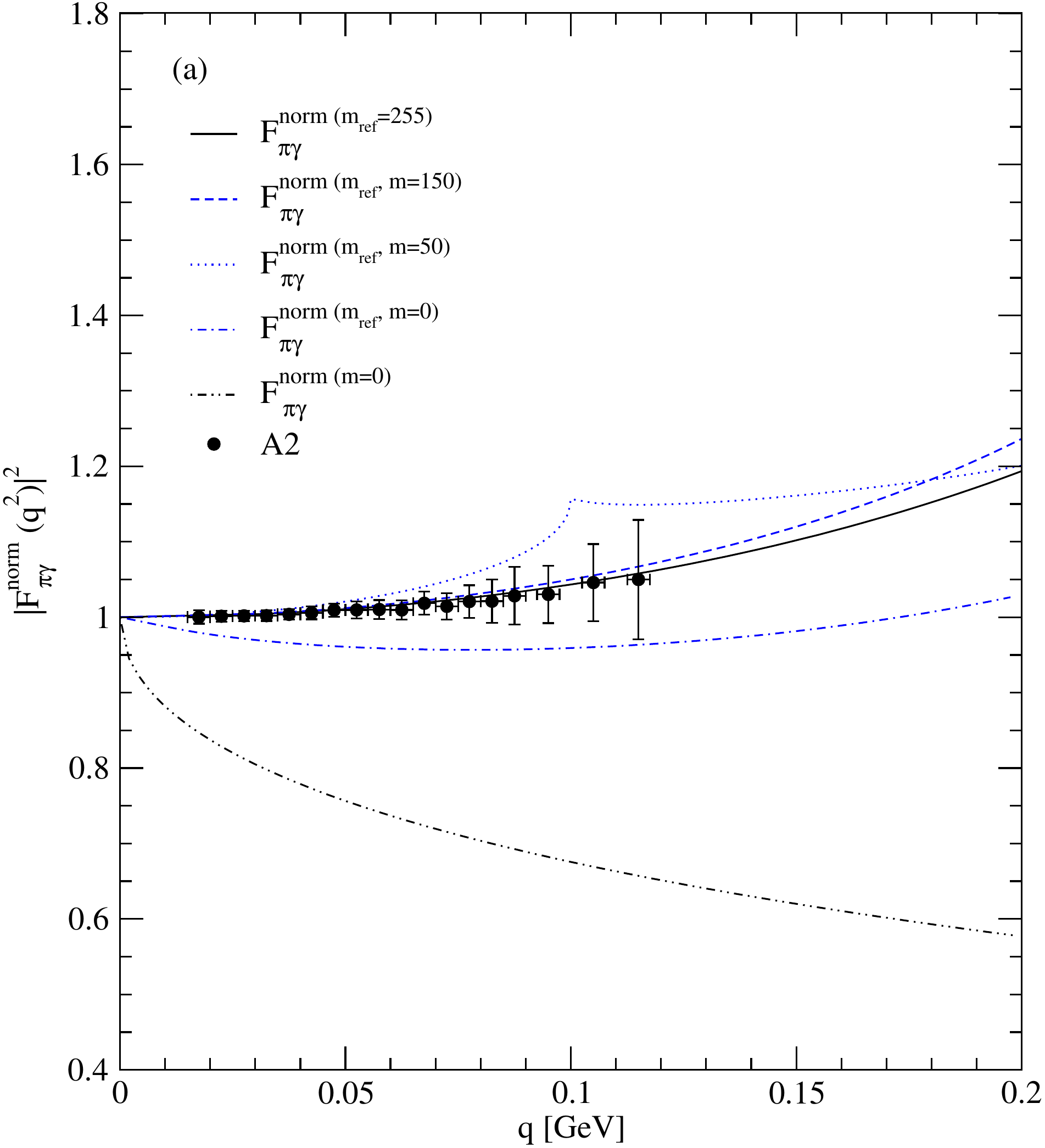}
\includegraphics[width=7cm, height=7cm]{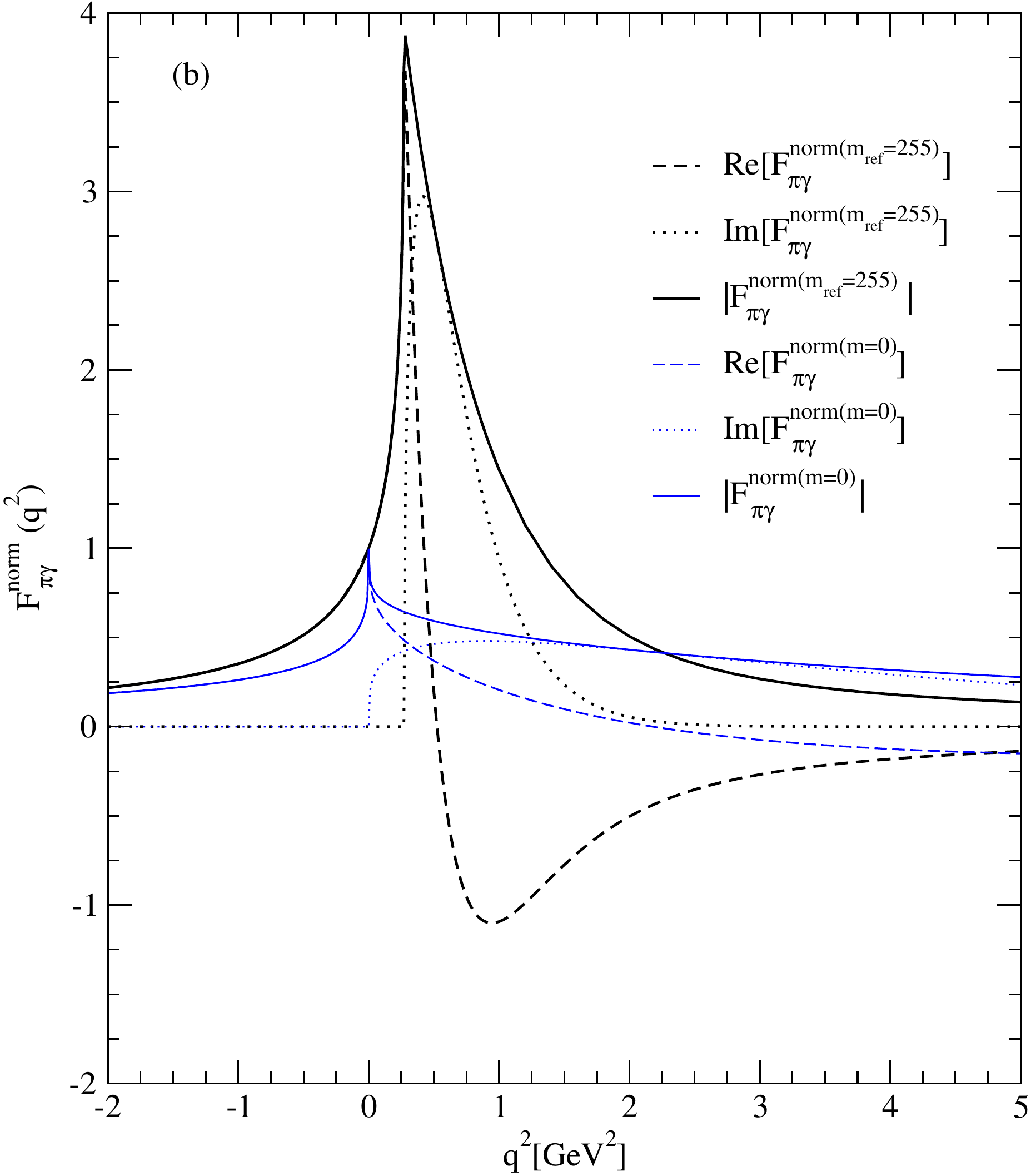}
\caption{\label{fig6} Predictions of normalized $F^{\rm norm}_{\pi\gamma}(q^2)=F_{\pi\gamma}(q^2)/F_{\pi\gamma}(0)$ in timelike region:
(a) $|F^{\rm norm}_{\pi\gamma}(q^2)|^2$ 
for small timelike region $(0\leq q\leq 0.2)$ GeV
and  (b) $F^{\rm norm}_{\pi\gamma}(q^2)$  for both spacelike and timelike regions ($-2\leq q^2\leq 5$) GeV$^2$.
The experimental data are taken from~\cite{A2}.
}
\end{center}
\end{figure*}

We show in Fig.~\ref{fig5}(a) our predictions of 
$F_{\pi\gamma}(Q^2)$ for low- and intermediate-spacelike regions of $0\leq Q^2\leq 10$ GeV$^2$ 
 and compare
with the experimental data~\cite{Behrend:1990sr,Gronberg:1997fj,Uehara:2012ag,Aubert:2009mc}.
We also show in Fig.~\ref{fig5}(b) our predictions of 
$Q^2 F_{\pi\gamma}(Q^2)$ for $0\leq Q^2\leq 50$ GeV$^2$ including large spacelike regions and compare with the data as well as
the leading twist pQCD prediction $Q^2 F_{\pi\gamma}(Q^2)=\sqrt{2}f_\pi$
 (double-dash-dotted line). 
 The solid and double-dot-dashed lines represent the results $F^{(m_{\rm ref}=255)}_{\pi\gamma}(Q^2)$
 and $F^{(m=0)}_{\pi\gamma}(Q^2)$
 obtained from  $m_{\rm ref}=255$ MeV  with $P_{m_{\rm ref}}=1$ and the chiral limit ($M_\pi,m\to0$) with $P_{m}=0.078$, respectively.
 The dashed, dotted, and dot-dashed lines represent the results 
 $F^{(m_{\rm ref}, m=150)}_{\pi\gamma}(Q^2)$, $F^{(m_{\rm ref}, m=50)}_{\pi\gamma}(Q^2)$, and $F^{(m_{\rm ref}, m=0)}_{\pi\gamma}(Q^2)$ 
obtained from the combination of the two different quark mass degrees of freedom
($m_{\rm ref}$ and $m=150$) MeV with the renormalized probability ${\tilde P}_{m}\approx 0.23$,  ($m_{\rm ref}$ and $m=50$) MeV  with ${\tilde P}_{m}\approx 0.13$, 
as well as ($m_{\rm ref}$ and $m=0$) MeV  with ${\tilde P}_{m}\approx 0.072$, respectively.
 We should note that all the results obtained from the parameter sets  in Table~\ref{t1} as well as their mixing
 satisfy the nonperturbative ABJ anomaly.
 While $F^{(m_{\rm ref})}_{\pi\gamma}(Q^2)$ agrees with the data for low $Q^2$ region ($Q^2<2$ GeV$^2$)
and shows the scaling behavior in the region above $Q^2\geq 10$ GeV$^2$,
the value of  $Q^2 F^{(m_{\rm ref})}_{\pi\gamma}(Q^2)$ as $Q^2\to\infty$ explains only about $80\%$ of the pQCD result. 
Our result $F^{(m=0)}_{\pi\gamma}(Q^2)$ obtained from the exact chiral limit (double-dot-dashed line) shows a disagreement with the experimental data
for low $Q^2(< 3$ GeV$^2)$ region. The $Q^2 F^{(m=0)}_{\pi\gamma}(Q^2)$ exceeds the  pQCD result for $Q^2 > 10$ GeV$^2$ and 
shows a consistency with the data from $BABAR$~\cite{Aubert:2009mc} for the intermediate region of $4\leq Q^2\leq 14$ GeV$^2$
although its mild rising behavior is however not enough to fit the data from $BABAR$ for the higher $Q^2$ region.

 The results of combining the two different quark mass degrees of freedom, i.e. $F^{(m_{\rm ref}, m=150)}_{\pi\gamma}(Q^2)$, $F^{(m_{\rm ref}, m=50)}_{\pi\gamma}(Q^2)$, 
 and $F^{(m_{\rm ref}, m=0)}_{\pi\gamma}(Q^2)$,
 are not much different from $F^{(m_{\rm ref})}_{\pi\gamma}(Q^2)$ for low- and intermediate-$Q^2$ regions
 but shows better agreement with the pQCD result in high $Q^2$ region accounting
 $93\%$ of the pQCD result. 
 While we have noticed that
$Q^2 F_{\pi\gamma}(Q^2)$ obtained from the quark mass in the region $0\leq m\leq 150$ MeV
 exceeds the pQCD result for $10\leq Q^2 \leq 20$ GeV$^2$ region by itself,
 it is interesting to see that the results of combining the quark mass degrees of freedom with $m_{\rm ref}$, i.e.
 $Q^2 F^{(m_{\rm ref}, m)}_{\pi\gamma}(Q^2)$, approach the asymptotic result only from
 below as shown in Fig.~\ref{fig5}(b).
 Effectively, our results obtained from the combination of the quark mass degrees of freedom 
 show a consistency with the data from Belle~\cite{Uehara:2012ag}
 rather than the $BABAR$ data~\cite{Aubert:2009mc}.
Our results for $F_{\pi\gamma}(0)$ are comparable with the simple LF holographic QCD model~\cite{AdS4} with a twist-2 valence pion state
in which it requires $P_{Q{\bar Q}}=0.5$ to reproduce $F^{\rm Exp}_{\pi\gamma}(0)$.
It may be also noteworthy that 
our previous LFQM prediction~\cite{Choi:2017zxn} using the spin-orbit structure given by Eq.~(\ref{eq9}) is
very close to the pQCD result.  

From the results of $F_{\pi\gamma}(Q^2)$ and $Q^2 F_{\pi\gamma}(Q^2)$ shown in Fig.~\ref{fig5}, we may summarize our findings as follows:
(1) For low- and intermediate-$Q^2$ region $(0\leq Q^2\leq 10)$ GeV$^2$ as shown in Fig~\ref{fig5}(a),
we find that the nonzero quark mass results are in better agreement with the data than 
the result in the chiral limit.
As the constituent quark mass decreases from the reference point $m_{\rm ref} =255$ MeV, the reduction of the probability
$P_{Q{\bar Q}}$ is necessary to agree with the experimental data.
These results indicate that the constituent quark picture ($2 m> M_\pi $) is very effective and important in describing $F_{\pi\gamma}(Q^2)$
in the low energy regime but the quark mass evolution seems inevitable as $Q^2$ grows.
(2) As the quark mass evolves from the constituent  to current quark masses, 
the probability $P_{Q{\bar Q}}$ finding the valence $Q{\bar Q}$ component inside the pion
also needs to be reduced accordingly.  This indicates that the higher Fock states contribute more as the quark mass decreases.

We show in Fig.~\ref{fig6}(a) the timelike ($q^2>0$) behavior of the normalized
$F^{\rm norm}_{\pi\gamma}(q^2)=F_{\pi\gamma}(q^2)/F_{\pi\gamma}(0)$, 
i.e. $|F^{\rm norm}_{\pi\gamma}(q^2)|^2$
as a function of $q$ for small $q$ ($0\leq q\leq 0.2$ GeV) region compared with the experimental data for the Dalitz decay
$\pi^0\to e^+e^-\gamma$ measured from A2 Collaboration~\cite{A2}. The same line codes are used as in Fig.~\ref{fig5}.
As discussed in Sec.~\ref{sec:III}, our result
for $F_{\pi\gamma}(q^2)$ in timelike region is obtained from the direct timelike region $(q^+=P^+)$ calculation without resorting to the analytic continuation
from spacelike $Q^2$ to the timelike $q^2$ in contrast to the case of the $q^+=0$ frame calculation.
Figure~\ref{fig6}(b) exhibits the sample results of $F^{\rm norm}_{\pi\gamma}(q^2)$ 
for both spacelike and timelike region ($-2\leq q^2\leq 5$ GeV$^2$)
obtained from
$m_{\rm ref}$  (black thick lines) and the exact chiral limit (blue thin lines), in which we separate
the real ${\rm Re}[ F_{\pi\gamma}(q^2)]$ (dashed lines) and imaginary 
${\rm Im}[ F_{\pi\gamma}(q^2)]$  (dotted lines) parts from the modulus
$|F_{\pi\gamma}(q^2)|$ (solid lines).
We should note that our direct results 
of the form factor $F_{\pi\gamma}(q^2)= {\rm Re}\; F_{\pi\gamma}(q^2) + i {\rm Im}\;F_{\pi\gamma}(q^2)$ 
are in complete agreement with those obtained from the dispersion
relations as we have explicitly shown in~\cite{Choi:2017zxn}.
This assures the validity of our numerical calculation both in the spacelike and timelike regions. 

In our model calculation for the timelike region, the imaginary part starts from the threshold, 
$q^2_{\rm th} = (m_Q + m_{\bar Q})^2=4 m^2$ and the modulus
of the TFF reaches maximum near threshold and decreases after the threshold. 
Because of this, $|F^{{\rm norm} (m_{\rm ref})}_{\pi\gamma}(q^2)|^2$  (solid line)
and $|F^{{\rm norm} (m_{\rm ref}, m=150)}_{\pi\gamma}(q^2)|^2$ (dashed line) in Fig.~\ref{fig6}(a)
represent the result including only the real part 
since the thresholds $q_{\rm th}$ for those quark masses (i.e. $m=255$ and 150 MeV) are greater than the maximum
$q$ value shown in Fig.~\ref{fig6}(a).  Both show an excellent agreement with the experimental data.
On the other hand, $|F^{{\rm norm} (m=0)}_{\pi\gamma}(q^2)|^2$ obtained from the chiral limit (double-dot-dashed line) 
represents the modulus including both real and imaginary parts but
reaches its maximum (see Eq.~(\ref{eq21})) at $q^2=0$ and decreases just after that. As a result,  the chiral limit prediction
in the timelike region shows an apparent disagreement
with the experimental data.
Likewise, $|F^{{\rm norm}(m_{\rm ref},m=50)}_{\pi\gamma}(q^2)|^2$ (dotted line) and $|F^{{\rm norm}(m_{\rm ref},m=0)}_{\pi\gamma}(q^2)|^2$ (dot-dashed line)
represent the modulus including both real and imaginary parts, but disagree with the data.
From the analysis of pion TFF in both spacelike and timelike regions, we find that the constituent quark picture is definitely necessary to describe the
low energy behavior correctly. While the form factors obtained from the mixing of the different quark mass eigenstates are not much different from each other in the spacelike region,
their predictions for the timelike region are very different due to the resonance feature in the timelike region. Therefore,
the analysis of $F_{\pi\gamma}(q^2)$ in timelike region plays a critical role in constraining theoretical models. 

In association with the experimental data for the Dalitz decay
$\pi^0\to e^+e^-\gamma$,  due to the smallness of the lightest $e^+e^-$ invariant mass $m_{ee}=q$, 
the normalized TFF is typically parametrized as~\cite{PDG2018,A2}
\be\label{eq26}
F^{\rm norm}_{\pi\gamma}(q^2) = 1 + a_\pi \frac{q^2}{m^2_{\pi^0}},
\ee
where the parameter $a_\pi$ corresponds to the slope of the TFF at $q^2=0$. 
As shown in Fig.~\ref{fig6}(a), our results for $F^{\rm norm}_{\pi\gamma}(q^2)$ obtained from
the two parameter sets, i.e. $m_{\rm ref}$ and the mixture of $m_{\rm ref}$ and $m=150$ MeV,
show a good agreement with the A2 data. 
Our results for $a_\pi$ obtained from $F^{{\rm norm} (m_{\rm ref})}_{\pi\gamma}(q^2)$ and $F^{{\rm norm} (m_{\rm ref}, m=150)}_{\pi\gamma}(q^2)$
are obtained as $a_\pi = 0.038$ and 0.043, respectively. On the other hand, in our previous LFQM~\cite{Choi:2017zxn}
analysis using the spin-orbit structure given by Eq.~(\ref{eq9}), we obtained $a^{\rm LFQM}_\pi = 0.036$.
Our result obtained from $m_{\rm ref}$ and the previous LFQM result~\cite{Choi:2007yu} are in closer good agreement with 
the current world average $a_\pi = 0.032 \pm 0.004$~\cite{PDG2018}  
and the two recent experimental data extracted from the $\pi^0\to e^+e^-\gamma$ decay, 
$a_\pi=0.030\pm 0.010$ from A2~\cite{A2}
and $a_\pi=0.0368\pm 0.0057$ from NA62~\cite{NA60}. This again indicates that the constituent quark degrees of freedom (i.e. $m\geq 200$ MeV) 
rather than the current quark degrees of freedom is much better in describing  $F_{\pi\gamma}(q^2)$ for
small timelike region. The timelike data going beyond the Dalitz decay can provide further constraints on theoretical understanding of the effective quark degrees of freedom.

\begin{figure}
\includegraphics[width=7cm, height=7cm]{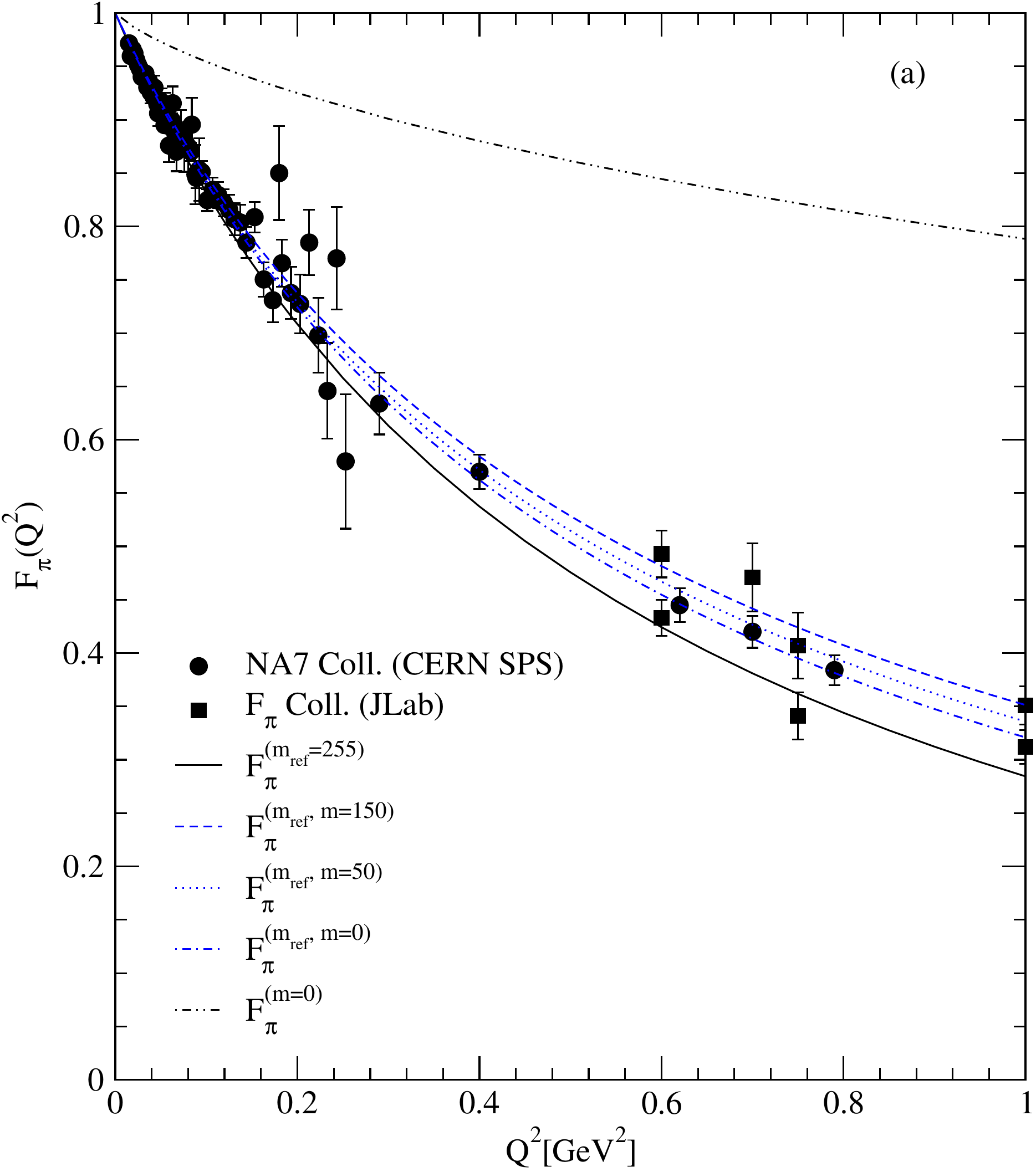}
\includegraphics[width=7cm, height=7cm]{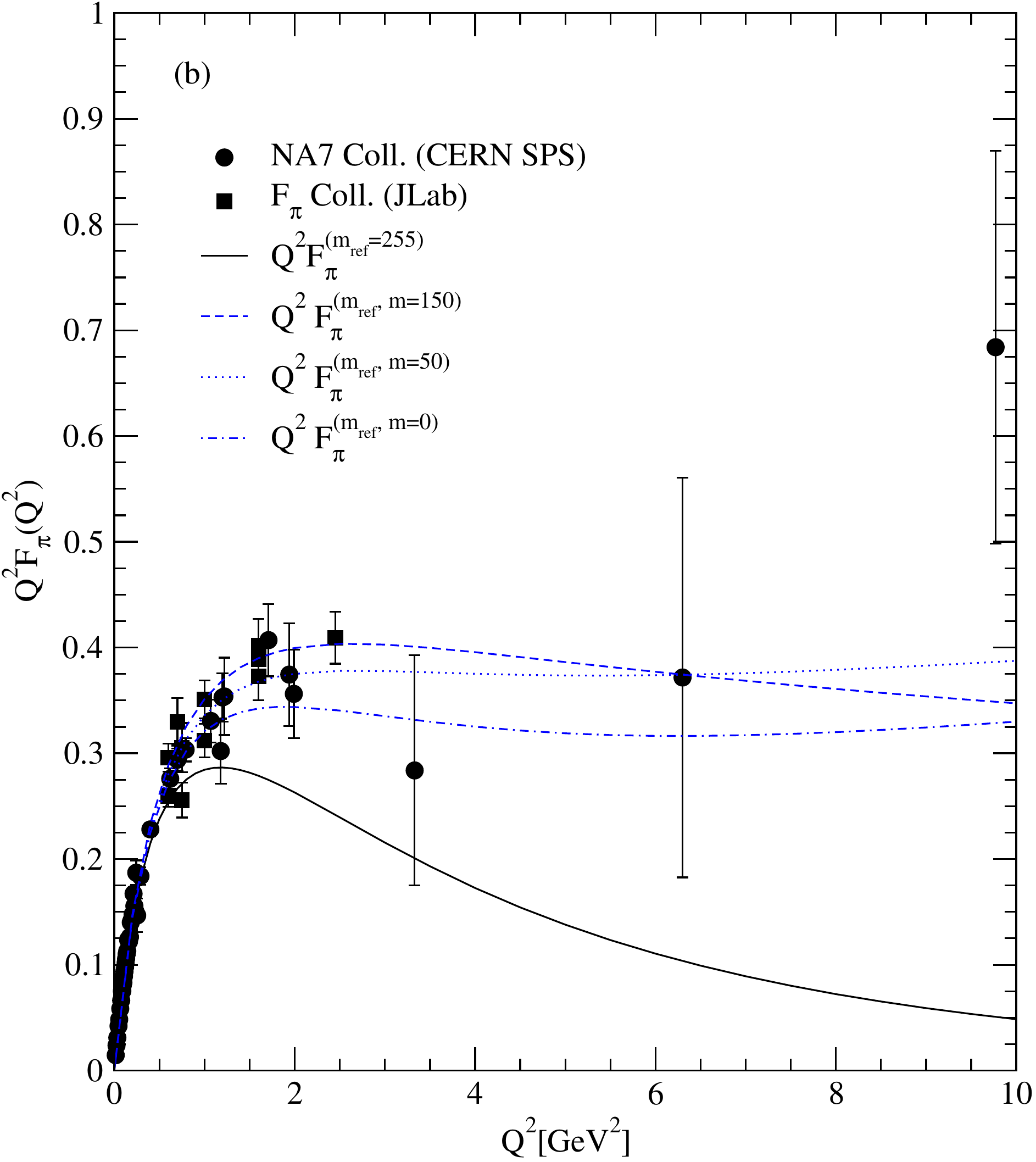}
\caption{\label{fig7}Predictions of (a) $F_{\pi}(Q^2)$ for small $Q^2$ $(0\leq Q^2\leq 1$ GeV$^2$) region and 
(b) $Q^2 F_\pi(Q^2)$ for the larger $Q^2$ ($0\leq Q^2\leq$ 10 GeV$^2$) region. The same line codes are used as in Fig.~\ref{fig5} and the data are taken from~\cite{Amen1,Volmer,Tade,Horn,Huber}.}
\end{figure}

Figure~\ref{fig7} shows the pion electromagnetic form factor, i.e. $F_\pi(Q^2)$  in Fig.~\ref{fig7}(a) for small $Q^2$ $(0\leq Q^2\leq 1$ GeV$^2$) region 
and $Q^2 F_\pi(Q^2)$ in Fig.~\ref{fig7}(b) for the larger $Q^2$ ($0\leq Q^2\leq$ 10 GeV$^2$) region.
We compare our results
with the experimental data~\cite{Amen1,Volmer,Tade,Horn,Huber}. 
The results $F^{(m_{\rm ref})}_{\pi}(Q^2)$ (solid line),  
$F^{(m_{\rm ref}, m=150)}_{\pi}(Q^2)$ (dashed line), $F^{(m_{\rm ref}, m=50)}_{\pi}(Q^2)$ (dotted line), and $F^{(m_{\rm ref}, m=0)}_{\pi}(Q^2)$ (dot-dashed line)
are in good agreement with the experimental 
data~\cite{Amen1,Volmer,Tade,Horn,Huber} for small $Q^2$ $(0\leq Q^2\leq 1$ GeV$^2$) region
as one can see from the plot of $F_\pi(Q^2)$ in Fig.~\ref{fig7}(a),
while the chiral limit result $F^{(m=0)}_{\pi}(Q^2)$ (double-dot-dashed line) severely deviates from the data
as one may have expected from the previous analysis of $F_{\pi\gamma}(Q^2)$. 
Our predictions of the pion charge radius $r_\pi \equiv \la r^2_\pi\ra^{1/2}$
obtained from  $F^{(m_{\rm ref})}_{\pi}$,  
$F^{(m_{\rm ref}, m=150)}_{\pi}$, $F^{(m_{\rm ref}, m=50)}_{\pi}$, and $F^{(m_{\rm ref}, m=0)}_{\pi}$
are given by $r_\pi = (0.683, 0.657, 0.677, 0.679)$ fm, respectively. Those four results show
a good agreement with  the most recent value
quoted by Particle Data Group~\cite{PDG2018}, $r^{\rm PDG}_\pi = (0.672\pm 0.008)$ fm.
These results may also be compared with the result $r^{\rm LFQM}_\pi = 0.651$ fm
obtained from the spin-orbit structure given by Eq.~(\ref{eq9}) in our previous LFQM analysis~\cite{Choi:2007yu}.
For the plots of $Q^2 F_\pi(Q^2)$ up to $Q^2=10$ GeV$^2$, 
the results of $Q^2 F^{(m_{\rm ref}, m=150)}_{\pi}(Q^2)$ (dashed line), 
$Q^2 F^{(m_{\rm ref}, m=50)}_{\pi}(Q^2)$ (dotted line), and $Q^2 F^{(m_{\rm ref}, m=0)}_{\pi}(Q^2)$ (dot-dashed line) appear reasonably
consistent with the current available experimental data, 
while the result $Q^2 F^{(m_{\rm ref})}_{\pi}(Q^2)$ reaches its maximum around $Q^2\simeq 1.2$ GeV and drops rather steeply after passing the maximum value.
Indeed, all the results for $Q^2 F_{\pi}(Q^2)$ 
drop after reaching their maximum values but with different rates.  
The result of combining the constituent quark mass degrees of freedom $Q^2 F^{(m_{\rm ref}, m=150)}_{\pi}$
provides an improved description over the result of $Q^2 F^{(m_{\rm ref})}_{\pi}$
without any mixing for the broader $Q^2$ range, 
which again indicates the vitality of quark mass evolution as $Q^2$ gets larger. We note however that 
the mixture of the constituent quark mass degrees of freedom with the current quark mass degrees of freedom as shown in $Q^2 F^{(m_{\rm ref}, m=50)}_{\pi}(Q^2)$ and $Q^2 F^{(m_{\rm ref}, m=0)}_{\pi}(Q^2)$ provides characteristically different
scaling behaviors compare to the typical high $Q^2$ behavior exhibited in the constituent quark quark picture results $Q^2 F^{(m_{\rm ref})}_{\pi}(Q^2)$ and $Q^2 F^{(m_{\rm ref}, m=150)}_{\pi}(Q^2)$.
We anticipate that the 12 GeV upgraded Jefferson Lab would provide much more
detailed and accurate data of the pion form factor for the larger $Q^2$ range. 
This would help us in coming up with the more realistic quark mass evolution analysis beyond this first order
approximation.

\section{Summary and Conclusions}
As the chiral anomaly~\cite{ABJ} is the key to understand the $\pi^0 \to \gamma\gamma$ decay rate
resolving the issue with the Sutherland-Veltman theorem~\cite{Sutherland-Veltman}, 
we attempt to include the axial-vector coupling in addition to the pseudoscalar coupling in our LFQM for the pion to explore a well-defined chiral limit still providing a good description of the pion electromagnetic and transition form factors~\cite{Choi:1997iq,Choi:2007yu,Choi:2017zxn}.
We thus took the spin-orbit vertex structure given by 
$\Gamma_\pi=(A _\pi + B_\pi {/\!\!\!\!\! P})\gamma_5$ versatile enough to explore the chiral limit  
and described the pion properties such as $f_\pi, F_{\pi\gamma}(Q^2)$ and $F_{\pi}(Q^2)$
depending on the variation of the quark mass in a  self-consistent manner.

We find that the chiral anomaly plays a critical role in constraining the model parameters.
The negativity of the axial-vector coupling, i.e. $B_\pi < 0$, appears essential to dictate 
the model independence of the ratio $\phi^{\rm chiral}_{\pi} (x)/ f^{\rm chiral}_\pi$
and the consistency with the AdS/CFT prediction of asymptotic DA.
Our chiral limit result for twist-2 pion DA given by Eq.~(\ref{eq16}) is exactly the same as the 
AdS/CFT prediction of the asymptotic DA~\cite{AdS1,AdS2,AdS3},
indicating also a significant higher Fock-state contribution in the chiral limit.
We also note that the analysis of $F_{\pi\gamma}(q^2)$ in timelike region plays a critical role 
in constraining theoretical model parameters. While the form factors obtained from the mixing of the different quark mass eigenstates are not much different from each other in the spacelike region,
their predictions for the timelike region are very different due to the resonance feature in the timelike region.

While the small probability of the lowest Fock-state such as $P_{Q{\bar Q}}<0.1$ in the chiral limit implies 
a significant higher Fock-state contribution, our numerical results in Sec.~\ref{sec:IV} indicate 
that $P_{Q{\bar Q}}$ increases as the quark mass increases. It is interesting to note that 
the merge of the parameter sets between the current quark picture and the constituent picture
occurs both in Fig.~\ref{fig3}(a) and Fig.~\ref{fig3}(b) as $P_{Q{\bar Q}}$ decreases. 
These results seem to reflect a nontrivial dynamic saturation process of 
the LF Fock-state expansion as the current quark degrees of freedom get amalgamated together  
to form the constituent quark degrees of freedom.

We may discuss the amalgamation of the current quarks forming the constituent quark degrees of freedom 
from the perspective of the vacuum fluctuation consistent with the chiral symmetry of QCD.
While the constituent degrees of freedom in our LFQM get dressed by the light-front zero-mode cloud, they satisfy the chiral symmetry consistent with the QCD. 
The correlation between the quark mass and the nontrivial QCD vacuum effect 
is on par with the trade-off between the complicated nontrivial vacuum and the effective 
constituent quark degrees of freedom.
Our results indicate that the constituent quark picture ($2 m> M_\pi $) is very effective and important in describing both $F_{\pi\gamma}(Q^2)$ and $F_\pi(Q^2)$ in the low energy regime but the quark mass evolution seems inevitable as $Q^2$ grows. More elaborate analysis including the quark mass evolution effect deserves 
further consideration. One may also explore the spectroscopic analysis including the pseudoscalar and vector meson nonets beyond the pion.

\label{sec:V}

\appendix*
\section{SPIN-ORBIT WAVE FUNCTIONS
$\chi(x,{\bf k}_{\perp})$ }
The constituent quarks  can be described by Dirac spinors
$u_\lam(k)$ and $v_\lam(k)$ satisfying the
Dirac equation
\be\label{Ap1}
  (/\!\!\!\! k-m) u_\lam(k)=0,\;\;\;
  (/\!\!\!\! k+m) \upsilon_\lam(k)=0, 
\ee
where $/\!\!\!\! k=k_\mu \gamma^\mu$.
It is instructive to use the appropriate basis of Dirac spinors~\cite{Jaus91}:
\be\label{Ap2}
u_\lam(k) 
    = \frac{1}{\sqrt{k^+}}(/\!\!\!\! k+m)u(\lam),\; \;
    \upsilon_\lam(k)
    = \frac{1}{\sqrt{k^+}}(/\!\!\!\! k-m) \upsilon(\lam), 
\ee
\be\label{Ap3}
u\left(\frac{1}{2}\right)\:
  =\left(
\begin{array}{c}
       1 \\
       0 \\
       0 \\
       0
\end{array}\; \right),
\; \; u\left(-\frac{1}{2}\right)
  =\left(
     \begin{array}{c}
       0 \\
       0 \\
       0 \\
       1
     \end{array} \right), 
\ee 
and $\upsilon(\lambda)=u(-\lam)$.
In this basis the $\gamma$ matrices are represented by
\be \label{Ap4}
\gamma^0\: = 
\: \left(
\begin{array}{cc}
0 & I \\
I & 0
\end{array} \; \right), 
\;\gamma^i = \left(
\begin{array}{cc}
0 & \sigma^i \\
-\sigma^i & 0
\end{array} \; \; \right),
\ee 
where $I$ is the $2\times 2$ unit matrix and $\sigma^{i}$ are 
Pauli matrices defined as 
\be\label{Ap5}
\sigma^1\:= \left(
\begin{array}{cc}
0 & 1 \\ 1 & 0
\end{array}\; \right), 
\;\sigma^2= \left(
\begin{array}{cc}
0 & -i \\ i & 0
\end{array}\; \right),
\; \sigma^3= \left(
\begin{array}{cc}
1 & 0 \\ 0 & -1
\end{array}\; \right). 
\ee
Using the $\gamma$ matrices  $\gamma^{\pm}\equiv\gamma^0\pm\gamma^3$
and $\gamma^5\equiv i\gamma^0\gamma^1\gamma^2\gamma^3$,
and $/\!\!\!\! k =\frac{1}{2} (k^+\gamma^- + k^-\gamma^+) -{\bf\gamma}_\perp\cdot{\bf k}_\perp$,
 the spinors $u_\lam(k)$ and $\upsilon_\lam(k)$
are obtained as
\be\label{Ap7}
u_{\uparrow} (k)\:=
\frac{1}{\sqrt{k^+}}
\left(
\begin{array}{c}
m \\ 0 \\ k^+ \\ k^R
\end{array}\; 
\right), \; 
\;u_{\downarrow}(k)=
\frac{1}{\sqrt{k^+}}
\left(
\begin{array}{c}
-k^L \\ k^+ \\ 0 \\ m
\end{array}\; \right),
\ee
\be\label{Ap8}
\upsilon_{\uparrow}(k)\:=
\frac{1}{\sqrt{k^+}}
\left(
\begin{array}{c}
        -k^L \\ k^+ \\ 0 \\ -m
      \end{array}
    \right),\; 
\; \upsilon_{\downarrow}(k)=
    \frac{1}{\sqrt{k^+}}
    \left(
      \begin{array}{c}
        -m \\ 0 \\ k^+ \\ k^R
      \end{array}\;
    \right).
\ee
The normalization is  $\bar{u}_\lam(k)u_\lam(k)=-{\bar \upsilon}_\lam(k)\upsilon_\lam(k)=2m$
and $k^R$ and $k^L$ are defined as
$k^R\equiv k^1+ik^2$ and $k^L\equiv k^1-ik^2$, respectively.

For a pion with four momentum $P$ and mass $M_\pi$, the general spin structure may be given 
as $\chi_{\lam\bar{\lam}}={\cal N} {\bar u}_{\lam 1}(k_1)(M_\pi+ B_\pi {/\!\!\!\!\! P})\gamma_5 \upsilon_{\lam_2}(k_2)$,
which satisfies the normalization 
$\sum_{\lam_i} \chi^{\dagger}_{\lam_1\lam_2}\chi_{\lam_1 \lam_2}=1$.

Then, the operator $\Gamma_\pi =(M_\pi + B_\pi {/\!\!\!\!\! P})\gamma_5$ is given by

\be\label{Ap11}
\Gamma_\pi =\left(
\begin{array}{cccc} 
             -M_\pi & 0 & B_\pi P^- & 0 \\
             0 & -M_\pi & 0 & B_\pi P^+ \\
             -B_\pi P^+ & 0 & M_\pi & 0 \\
             0 &  -B_\pi P^- & 0 & M_\pi
\end{array}\; \right), 
\ee
and 
 \bea\label{Ap13}
&&(1) \; \chi^\pi_{\uparrow\uparrow}={\cal N}
\frac{-k^L_1}{\sqrt{x_1x_2}} ( M_\pi + 2B_\pi m), 
\nonumber\\
&&(2) \; \chi^\pi_{\downarrow\uparrow}={\cal N}
\frac{-1}{\sqrt{x_1x_2}} \{ m M_\pi - B_\pi ( {\bf k}_\perp^2 - m^2 - x_1 x_2 M_\pi^2)\},
\nonumber\\
&&(3) \; \chi^\pi_{\uparrow\downarrow}={\cal N}
\frac{1}{\sqrt{x_1x_2}} \{ m M_\pi - B_\pi ( {\bf k}_\perp^2 - m^2 - x_1 x_2 M_\pi^2)\},
\nonumber\\
&&(4) \;  \chi^\pi_{\downarrow\downarrow}={\cal N}
\frac{-k^R_1}{\sqrt{x_1x_2}} ( M_\pi + 2B_\pi m), 
\eea
where 
 \be\label{Ap14}
 P=\left(P^+, \frac{M^2}{P^+}, {\bf 0}_\perp \right),\;\;
 k_i=  \left(x_i P^+, \frac{{\bf k}^2_{i\perp} + m^2_i}{x_i P^+}, {\bf k}_{i\perp} \right).
 \ee

Thus, the normalized spin-orbit wave function for pion 
satisfying the unitary condition $\langle\chi_{\lam_1\lam_2}|\chi_{\lam_1 \lam_2}\rangle=1$
is given by
\be\label{Ap16}
\chi_{\lam_1\lam_2}(x,{\bf k}_{\perp}) = {\cal N}
\left(
\begin{array}{cc}
        -k^L{\cal M} & m{\cal M} + x_1x_2B_\pi \epsilon_B \\ 
       -m{\cal M} - x_1x_2 B_\pi \epsilon_B  & -k^R {\cal M}
      \end{array}
    \right),\;
\ee
where  ${\cal N}=  \frac{1}{\sqrt{2}\sqrt{ {\cal M}^2 \;{\bf k}^2_\perp + [m{\cal M} +  x_1 x_2 B_\pi \epsilon_B]^2}}$,
${\cal M}= M_\pi +  2B_\pi m$ and $\epsilon_B=M^2_\pi - M^2_0$.

\acknowledgments
H.-M. Choi was supported by the National Research Foundation of Korea (NRF)
(Grant No. NRF-2020R1F1A1067990). 
C.-R. Ji was supported in part by the US Department of Energy
(Grant No. DE-FG02-03ER41260).

\end{document}